\documentclass[twocolumn]{aastex63}

\usepackage{amsmath}
\usepackage[hyphens]{url}

\newcommand\teff{T_{\rm eff}}
\newcommand{\e}[1]{ \times 10^{#1}}
\newcommand{\wig}[1]{\mathrel{\hbox{\hbox to 0pt{%
          \lower.6ex\hbox{$\sim$}\hss}\raise.4ex\hbox{$#1$}}}}
\newcommand\sss{\scriptscriptstyle}

\newcommand\kb{k_{\sss\rm B}}

\shorttitle{Diffusion coefficients in white dwarfs}
\shortauthors{Heinonen et al.}

\begin{document}


\title{Diffusion coefficients in the envelopes of white dwarfs}


\author{R. A. Heinonen}
\affil{Dept. of Physics, University of California San Diego,  9500 Gilman Dr, La Jolla, CA 92093}

\author{D. Saumon}
\email{dsaumon@lanl.gov}

\author{J. Daligault}

\author{C. E. Starrett}
\affil{Los Alamos National Laboratory, PO Box 1663, Los Alamos, NM 87545}

\author{S. D. Baalrud}
\affil{Dept. of Physics and Astronomy, University of Iowa, Iowa City, IA, 52242}

\author{G. Fontaine}
\affil{D\'epartement de Physique, Universit\'e de Montr\'eal, Montr\'eal, QC H3C 3J7, Canada}




\begin{abstract}
The diffusion of elements is a key process in understanding the unusual surface composition of white dwarfs stars
and their spectral evolution.  The diffusion coefficients of
Paquette et al. (1986) have been widely used to model diffusion in white dwarfs.
We perform new calculations of the coefficients of inter-diffusion and 
ionic thermal diffusion with 1) a more advanced model that uses a recent modification of the calculation of 
the collision integrals that is more suitable for the partially ionized, partially degenerate and moderately
coupled plasma, and 2) classical molecular dynamics. The coefficients are evaluated for silicon and calcium 
in white dwarf envelopes of hydrogen and helium.  A comparison of our results with Paquette et al. shows that the latter 
systematically underestimates the coefficient of inter-diffusion yet provides
reliable estimates for the relatively weakly coupled plasmas found in nearly all types of stars as well as in white dwarfs
with hydrogen envelopes. In white dwarfs with cool helium envelopes ($\teff < 15000\,$K), the difference grows to more
than a factor of two.  We also explored the effect of the ionization model used to 
determine the charges of the ions and found that it can be a substantial source of discrepancy between different calculations.
Finally, we consider the relative diffusion time scales of Si and Ca in the context of the pollution of white dwarf photospheres
by accreted planetesimals and find factor of  $\wig> 3$ differences between calculations based on Paquette et al. and our model.
\end{abstract}





\section{Introduction}

White dwarfs (WDs) constitute the end stage of the evolution of the vast majority of stars and are quite common 
in the Galaxy.  With masses of about 0.6$\,$M$_\odot$ and very small radii of $\sim 10^9\,$cm, they have
high surface gravities of $\log g \,({\rm cm/s}^2) \sim 8$ and are characterized by exotic 
physical conditions such as central densities of up $\sim 10^6\,$g/cm$^3$ that are well above those of normal stars and far 
beyond the reach of current experimental capabilities. On the other hand, WDs are so common that
they are very well characterized observationally. They form a fascinating class of stars for the application of theories of
dense matter.

In particular, the spectra of WDs indicate that their atmospheric compositions are unlike those of other
stars, being composed of pure hydrogen or helium with 25-50\% of all WDs showing a small 
amount of ``pollution'' from a few
heavier elements \citep{koester14}.  \citet{schatzman58} estimated that their high surface  gravity would lead to
a gravitational settling of the heavier elements by diffusion on time scales of $\sim 10^6\,$ years, \footnote{Modern calculations give diffusion time scales of
days to millions of years at the bottom of the convection zone.} 
which explained the nearly pure composition of most WD atmospheres and established the importance of diffusion in WDs. 
Subsequent studies of the role of diffusion in WD envelopes and atmospheres, 
combined with other processes such as accretion, convective dredge up, and radiative forces have led to an
understanding of the various spectral types of WDs as well as their {\it spectral evolution} (see \cite{blouin19a} for a summary).


The astrophysical interpretation of a large body of WD observations requires a knowledge of the coefficient of inter-diffusion
in dense plasmas. This is a challenging endeavor as the plasma can be partially ionized,
with partially degenerate electrons, and weakly to strongly coupled ions. Historically, models of diffusion in WDs used coefficients
based on the solution of the Bolztmann equation of the kinetic theory of gasses, as obtained by the Chapman and Enskog method  \citep{cc70} or
the approach of \cite{burgers69}.
In both approaches, transport coefficients are expressed in terms of binary collision integrals, which is appropriate for
dilute gasses \citep{ac60, michaud70, michaud76, fm79, im85}. In low density stellar plasmas, the interaction responsible
for the ion collisions can be described as a Coulomb potential with an appropriate  choice of long-range cutoff.
A considerable advance in physical realism was achieved by \cite{liboff59, mason67, muchmore84} and \citet{paquette86a} who introduced
the static screened Coulomb potential (also called Yukawa or Debye-H\"uckel) in the collision integrals with a screening 
length that accounts for electron and ion screening as well as the 
strong screening limit.  The tabulated collisions integrals of \citet{paquette86a} have been widely applied in stellar models and to
WD models in particular (see, for example, \cite{pelletier86, dupuis92, althaus00, koester09}). A slightly modified physical model
has been applied in more recent evaluations of the collision integrals \citep{fbdt15, stanton16}.

The last three decades have seen considerable development in the theory of dense plasmas,  advanced 
computer simulation methods, and computational capabilities that justify new calculations of transport coefficients in
regimes relevant to WDs. In this paper, we calculate new
diffusion coefficients with a more realistic physical model that relaxes several of the key assumptions of 
\citet{paquette86a}.  Specifically, we present diffusion coefficients of calcium  and silicon ions in plasmas of H and He,
at the conditions found at the bottom of the superficial convection zone of WDs and compare with the results of \citet{paquette86a}
and others.

The paper is structured as follows.  We first outline in section \ref{sec:methods_diff} the three methods that we apply to the calculation of
coefficients of diffusion and discuss how they relate to each other as well as their merits and limitations. The details of our calculations of the
inter-diffusion and ionic thermal diffusion coefficients are given in Section \ref{sec:results_diff}, where we compare the results from various methods.
One of the most interesting applications of diffusion in white dwarfs is the combination of accretion and diffusion of metals that explains their presence 
in the photospheres of DZ and DAZ stars.  This process is characterized by the diffusion time scale at the bottom of the convection zone. In section 
\ref{sec:results_time} we compare the diffusion time scales of Si and Ca predicted by three models.
Section \ref{sec:conclusion} summarizes our work and results and provides a broader perspective on modern calculations of transport coefficients in white dwarf stars.

\section{Modeling diffusion coefficients in dense plasmas}
\label{sec:methods_diff}

In most instances of diffusion in WDs, we are concerned with the inter-diffusion in a binary mixture where one of the
species is present as a trace, which is the focus of this paper\footnote{The methods described herein can be applied to the calculation of transport coefficients in plasmas of arbitrary 
mixtures of elements.}. For a binary mixture of species with number concentrations $x_1$ and $x_2$ ($x_1 + x_2 = 1$), arbitrary electron degeneracy, and in the limit where
$x_2 \ll 1$, the equation for the relative velocity of the species $w_{12}$ due to diffusion is 

\begin{align}
	w_{12}  = D_{12} \bigg[& -\frac{\partial \ln x_2}{\partial r} + 
	\Big( \frac{Z_2}{Z_1} A_1 - A_2 \Big) \frac{m_0 g}{\kb T}  \nonumber \\
	 & + \Big( \frac{Z_2}{Z_1} - 1\Big) \frac{\partial \ln P_i}{\partial r} +
        \alpha_{\sss T} \frac{\partial \ln T}{\partial r} \bigg]
\label{w12}
\end{align}
 
where $D_{12}$ is the coefficient of inter-diffusion, $r$ the radius inside the star, $A_i$ and $Z_i$ the atomic mass (in a.m.u.) 
and charge of ions of species $i$ ($i=1$ for the light background 
ion (H or He) and $i=2$ for the heavy ion), $P_i$ the {\it ionic} pressure, $T$ the temperature, $\kb$ the Boltzmann constant, $m_0$ the atomic mass unit, and
$\alpha_{\sss T}$ is the thermal diffusion factor \citep{pelletier86, bauer19}. The first term on the right hand side of Equation \ref{w12} is 
driven by concentration gradients and corresponds to ``ordinary'' chemical diffusion. The second and third terms describe barodiffusion, also known as gravitational 
settling in stars, 
caused by pressure gradients associated with the star's gravitational field and the induced electric field. This is generally the dominant term in white 
dwarf envelopes \citep{paquette86a}.  The last term is the 
contribution of thermal diffusion.  Equation \ref{w12} neglects the contribution of radiative forces which are negligible in the relatively cool 
white dwarfs we are considering here \citep{chayer95}. Furthermore, Equation \ref{w12} applies to a single trace ion of charge $Z_2$ but in general the trace element 
has a distribution of charge states, with each ion species described by a separate diffusion equation. An average diffusion velocity for a given element can be defined 
for such a multi-component diffusion problem but
in the following we make the simpler and common approximation of using a single diffusion equation of an element with an average {\it ion charge} \citep{dupuis92, koester09, bauer19}.
When $w_{12} > 0$ in Equation \ref{w12}, species 2 moves toward larger $r$ (toward the surface). While Equation \ref{w12} is valid for $x_2 \ll 1$,  the calculation of $D_{12}$
is not contingent on that limit. However, the examples presented in section \ref{sec:results_diff} are all for cases where species 2 is a trace.  
In plasmas, the coefficient of thermal diffusion $D_{\sss T} = D_{12} \alpha_{\sss T}$ is determined by collisions between ionic species and 
between ions and electrons and can be written as $\alpha_{\sss\rm T}=\alpha_{12} + \alpha_{1e} + \alpha_{2e}$ \citep{paquette86a}. In this paper, we consider only the ionic
term $\alpha_{12}$. The contributions of $\alpha_{1e}$ and $\alpha_{2e}$ can be comparable or even much larger than $\alpha_{12}$ \citep{paquette86a} and 
will be the subject of a future publication. 

There are several approaches to the calculation of $D_{12}$, three of which are compared here: 1) direct simulation
with molecular dynamics, 2) the model of \citet{paquette86a}, and 3) the effective potential theory which we have developed. We
first summarize the main features, advantages and drawbacks of each method.

\subsection{Molecular dynamics}

Classical molecular dynamics is a simulation method that, when applied to particles such as ions in a plasma, and {\it given an ion-ion
pair potential}, allows for the direct evaluation of ionic transport coefficients without any assumption beyond those implicit in the input pair potential.   
Applying the fluctuation-dissipation theorem,
it can be shown that the diffusion coefficients $D_{12}$ and $\alpha_{12}$ can be evaluated from a simulation of the system in equilibrium  --
there is no need to simulate a system with the external gradients that appear in Equation \ref{w12}.  In such a simulation, 
a large number of classical particles ($\sim 10^3$) is set in a cubic box, each  with an initial position and a velocity sampled from a Maxwell-Boltzmann distribution.  The total force
on each particle is summed from the pair interactions with all the other particles. The positions and velocities are advanced in time according
to Newton's second law. An infinite system is approximated by replicating the box in three dimensions with periodic boundary conditions.  After  a
period of equilibration to an imposed value of the temperature, the particles are allowed to move over a relatively long period of time.  
An analysis of
the positions, velocities along the trajectories of the particles can be performed to evaluate many physical properties of interest.  


In stellar astrophysics, we are generally concerned with diffusion between two ionic species. The microscopic definition of 
the coefficient of inter-diffusion between species 1 and 2 is obtained from a time auto-correlation function \citep{green54, kubo57, mcquarrie, haxhimali14}
\begin{equation}
     D_{12}= \frac{J}{3N x_1 x_2} \int_0^\infty \langle {\bf j}(t) \cdot {\bf j}(0) \rangle \, dt
     \label{diff_micro}
\end{equation}
where
\begin{equation}
	{\bf j}(t) = x_2 \sum_{i=1}^{N_1} {\bf v}_{1,i}(t) - x_1 \sum_{i=1}^{N_2} {\bf v}_{2,i}(t)
\end{equation}
is the net particle current in a system of $N_j$ particles of species $j$ of number fraction $x_j=N_j/N$, and $N=N_1+N_2$ is the total number of particles.
The velocity of particle $i$ of species 1 at time $t$ is ${\bf v}_{1,i}(t)$, which is extracted from the particle trajectories of the simulation.  
The factor $J$ is the so-called thermodynamic factor
\begin{equation}
	J = \frac{x_1 x_2}{\kb T} \frac{\partial \mu_1}{\partial x_1} \Bigg|_{P,T}
	\label{thermo_factor_1}
\end{equation}
where $\mu_1$ is the chemical potential of species 1.  This factor converts the diffusion coefficient
defined in terms of the gradient in the chemical potential (Maxwell-Stefan equation of diffusion) to the coefficient defined in terms of the gradient in the concentration 
(Fick's equation of diffusion).  In the limit of an ideal gas of neutral particles or for the diffusion of a trace element, $J=1$. The thermodynamic factor
can be evaluated from the equation of state (e.g. Equation \ref{thermo_factor_1}) or, as we have done, from the ionic structure factors $S_{ij}(k)$ of the plasma mixture 
\begin{equation}
	J^{-1} = x_2 S_{11}(0) + 2\sqrt{x_1 x_2} S_{12}(0) + x_1 S_{22}(0).
	\label{thermo_factor_2}
\end{equation}
The structure factors $S_{ij}(k)$ are the Fourier transforms of the pair distribution functions $g_{ij}(r)$ that
in turn describe the (normalized) radial density profiles of ions of species $j$ around an ion of species $i$. For an ideal gas, ions are spatially uncorrelated and $g_{ij}(r)=1$ but with increasing interactions the
ions become correlated, which is reflected in the structure of $g_{ij}(r)$.
The evaluation of equation \ref{diff_micro} is straightforward from the particle positions produced in a classical MD simulation, although care must be 
exercised to obtain well-converged results. 

The coefficient of {\it ionic} thermal diffusivity $D_{\sss T} = \alpha_{12}D_{12}$ can be evaluated in a similar fashion from classical MD simulations. In this case however, 
the statistical sampling of the trajectories is much less efficient and the resulting coefficient becomes very noisy. For this reason, we do not 
present molecular dynamics results for $\alpha_{12}$.


While MD is appealing for its conceptual simplicity and minimal set of assumptions, it does have several numerical drawbacks. Because the system being
simulated is relatively small compared to any macroscopic system, it is subject to statistical fluctuations that are particularly significant when
evaluating quantities such as $D_{12}$ and $\alpha_{12}$. They can be reduced at the cost of increasing the length of
the simulation and the number of particles.  As is often the case in WDs, when one species is present as a trace (e.g. $x_2 \ll 1$) the simulation box contains only a small number of particles
of the trace species, greatly increasing the statistical noise in $D_{12}$.   Highly asymmetric mixtures, with high mass or charge ratios between the ionic
components, are computationally more demanding because of the very different dynamical time scales and interaction forces of the two species \citep{ticknor16}.  Such mixtures 
are typical of white dwarfs where the background species is usually H or He and the diffusing species of interest has an atomic number $Z \sim 6 - 21$. 
Finally, weakly coupled systems can take a very long time to equilibrate as collisions 
are weak or infrequent because of low density and energy exchange between particles proceeds slowly.  
Nonetheless, with current high performance computers, very large and very long MD simulations can be performed to accurately evaluate $D_{12}$ for mixtures 
\citep{haxhimali14} but this is not practical to generate tables for astrophysical applications.

The values of $D_{12}$ evaluated with classical MD are of course only as reliable as the ion-ion potential that is provided as input to the simulation. In
dense plasmas, relatively simple potentials are often used such as a pure Coulomb interaction in the one component plasma and binary ionic mixture models 
\citep{hansen75, hansen79, bastea05, daligault12, shaffer17},  or of
a Yukawa form \citep{salin06,haxhimali14}. These model potentials represent the limits of a rigid electron background (no screening) and screening in the weakly coupled limit, respectively,
and can be accurate in the appropriate physical regimes. As we will see below, realistic self-consistent potentials for
dense plasmas can be obtained from an average atom model, without any assumption for its functional form. 

The most accurate approaches are {\it quantum} MD and
{\it orbital free} MD that do not require a ion-ion pair potential.  Instead, the simulation considers only classical nuclei and quantum electrons and almost always in the 
Born-Oppenheimer approximation where the kinetic degrees of freedom of the electrons are decoupled from those of the ions.  
The density of the quantum electron fluid in the 
simulation box is calculated by solving the Schr\"odinger equation or the Thomas-Fermi model\footnote{In Orbital-Free Molecular Dynamics (OFMD), the expensive calculation of  
the quantum mechanical wave functions is eliminated and the electrons are described in terms of the electron density only.  The semi-classical Thomas-Fermi model is the simplest and most commonly 
used orbital-free model.}, respectively, for each configuration of the nuclei at each time step. The force on each nucleus is the sum of
the forces from all the other nuclei and from the 3-dimensional density of the electron fluid.  Unfortunately, these methods are computationally very 
intensive, which severely limits the size and length of the simulations, but have nevertheless been applied to the computation of the self- and inter-diffusion
coefficients \citep{lambert06, kress11, danel12, rudd12, french12, jakse13, burakovsky13, meyer14, sjostrom15, ticknor15}.
Here, we use classical MD simulations to evaluate coefficients of inter-diffusion for the purpose of validating the results obtained with 
the \cite{paquette86a} model and the effective potential theory method that we have developed.

\subsection{Introduction to the kinetic theories of Paquette et al. and Effective Potential Theory}

A goal of kinetic theory is to express as accurately as possible formal relations such as Equation \ref{diff_micro} explicitly in terms of the 
interaction potentials between particles and in a form that makes their numerical evaluation straightforward in comparison with molecular dynamics simulations.
Models of this kind for the coefficient of ionic inter-diffusion in stars
are based on the expression derived from the solution of the Boltzmann kinetic equation \citep{burgers69, cc70}.
The resulting transport coefficients are expressed in terms of collision integrals involving scattering cross sections for isolated, binary collisions.
The application of this approach to plasmas is not straightforward.
Indeed, strictly speaking, the Boltzmann equation is only valid for gases of particles with short-range binary interactions that are dilute enough that 
the particle dynamics can be described as a succession of spatially localized and uncorrelated binary encounters.
This approximation does not directly apply to a plasma regardless of its density or its temperature, because the long-range nature of the Coulomb potential 
invalidates the assumption of spatially localized collisions and, as a consequence, leads to divergent collision integrals.
Yet, the difficulty can be remedied by considering that in a plasma, two charged particles never interact directly via the Coulomb potential 
because the presence of the surrounding particles screens the interaction, resulting in an effective binary interaction potential that is short-ranged.
When the latter is used in the Boltzmann equation in place of the bare Coulomb potential, the resulting transport coefficients remain finite.
The simplest form of effective interaction is a pure Coulomb potential with a long-range cutoff, usually chosen as the Debye screening length (Equation \ref{debye}).
In accordance with Boltzmann's theory, the effective potential approach is appropriate if the dynamics of charged particles can be described as a succession of 
binary collisions in this effective potential.  This is the case in weakly coupled plasmas, i.e. under density and temperature conditions that are such that 
the typical kinetic energies of particles is much larger than their typical mutual interactions.  Interestingly, in WD plasmas, electrons are generally 
weakly coupled to other electrons and to ions either because of the high temperature or the high quantum degeneracy.
Conversely, the ions typically remain non-degenerate and can become strongly coupled at low enough temperature.

The two kinetic theories used in this work, namely the model of Paquette et al. and the more advanced effective potential theory, rest upon two different 
models of the effective ion-ion potential designed to give accurate transport coefficients in the weak and the moderately coupled plasma regimes, respectively.
The Paquette et al. approach assumes that the effective pair interaction is a screened Coulomb potential, which accounts for screening by both electrons and ions in the 
weakly coupled limit. More generally, \cite{baalrud19} have shown that the postulate of an effective pair interaction in the Boltzmann equation can be rigorously derived from
a kinetic theory based on an expansion in terms of the departure of correlations from their equilibrium values. 
The derivation also yields the proper form of the potential in the Effective Potential Theory, which we apply in this study.


\subsection{Model of Paquette et al. (1986)}
\label{sec:paquette}

The strength of the coupling between two ions of charge $Z_1$ and $Z_2$ is conveniently quantified by the coupling parameter
\begin{equation}
  \Gamma = \frac{Z_1 Z_2 e^2}{a\kb T}\,, 
  \label{gamma}
 \end{equation}
which is the ratio of the Coulomb energy between two neighboring ions to their kinetic energy,  $a$ being the ion sphere radius, and $e$ the (positive) quantum of charge.
Ions are weakly coupled when $\Gamma \ll 1$.
In weakly coupled plasmas, the linear response approximation can be used to determine the effective interaction between two ions \citep{eliezer02}.
This yields the static screened Coulomb potential 
\begin{equation}
	V_{\sss\rm D}(r) = \frac{Z_1 Z_2 e^2}{r} e^{-r/ \lambda_{\sss D}} 
	\label{yukawa}
\end{equation}
where 
\begin{equation}
\lambda_{\sss D}= \bigg[ \lambda_i^{-2}+\lambda_e ^{-2} \bigg ] ^{-1/2}
   \label{debye}
\end{equation}
is the total screening length that accounts for the screening of the ion interactions by both 
ions and electrons, with $\lambda_i^2=\kb T/4\pi e^2\sum_i{n_i Z_i^2}$ the classical Debye-H{\"u}ckel screening length of ions and $\lambda_e$ the 
electron screening length.  The regime of validity of this approach can be extended to $\Gamma \sim 0.3$.
Beyond, the weak coupling approximation fails and the screening is no longer described by Equation \ref{debye}.
It was suggested on physical grounds that an approximate potential that extends the domain of validity of the screened potential (\ref{yukawa}) 
is obtained by replacing the screening length by the ion sphere radius $a$.
Thus the weak coupling limit and an approximation of the strong coupling limit can be obtained by choosing a screening length
\begin{equation}
    \lambda = \max\{\lambda_{\sss D}, a\}.
    \label{l_max}
\end{equation}
\citet{muchmore84}, \citet{paquette86a}, and \citet{brassard14} used the potential defined by Equations (\ref{yukawa}--\ref{l_max})
to evaluate the coefficients of inter-diffusion and thermal diffusion, using the classical limit for the electron screening length $\lambda_e^2= \kb T/4 \pi e^2 n_e$.
Recently, \cite{fbdt15} and \citet{stanton16} revised the work of \citet{paquette86a} by using the Thomas-Fermi screening length for the electrons
to account for electron degeneracy
\begin{equation}
	\lambda_{\sss\rm TF}^2 =  2\frac{I_{1/2}(\beta \mu_{\rm e})}{I_{-1/2}(\beta \mu_{\rm e})} \lambda_e^2,
	\label{TF_length}
\end{equation}
where $\beta=1/\kb T$, $\mu_{\rm e}$ is the chemical potential of the electrons and 
\begin{equation}
	I_n(\alpha)=\int_0^\infty \frac{x^n}{e^{x-\alpha}+1} \, dx
\end{equation}
is the Fermi integral of index $n$. Equation \ref{TF_length} reduces to the classical Debye-H\"uckel length in the non-degenerate limit. 
\footnote{\cite{stanton16} use a fit to the exact expression given by Equation \ref{TF_length}.}
Both of these works also implemented a smoother transition to the strongly coupled approximation by replacing 
Equation \ref{l_max} with
\begin{equation}
	\lambda=\frac{\lambda_{\sss\rm D}^5 + a^5}{\lambda_{\sss\rm D}^4 + a^4}
	\label{lambda_fbdt15}
\end{equation}
\citep{fbdt15} while \cite{stanton16} chose a more  physically motivated interpolation form
\begin{equation}
	\lambda=\Bigg( \frac{1}{\lambda_{\sss\rm TF}^2} + \frac{1}{\lambda_i^2 + a^2} \Bigg) ^{-1/2}.
	\label{lambda_sm16}
\end{equation}

The collision integrals involving the static screened potential must be evaluated numerically. However, a considerable advantage of this approach is that
they can be expressed in dimensionless form with the parameters of the specific mixture (masses and charges of the ions) factored out \citep{paquette86a}. Thus,
the dimensionless collision integrals for the static screened potential can be evaluated once and for all for any mixture.  On the other hand, 
for dimensional calculations of diffusion coefficients, a separate model is needed to provide the charges of the ions $Z_i$ that enter the pair potential.
The simplest approach is to estimate the ion charge with the Thomas-Fermi average atom model \citep{stanton16}.
The choice of ionization model is a source of uncertainty in the calculated values of $D_{12}$ \citep{bauer19} and transport coefficients in general \citep{graboske20}. 
We will return to this point in section \ref{sec:results_diff}.

Since their publication, the fits of the collision integrals of \citet{paquette86a} have been used in nearly all models of diffusion in white 
dwarfs stars, and in numerous calculations of diffusion in stars in general. Because of their prevalence in stellar astrophysics, they
represent a standard for comparison with our own calculations of $D_{12}$ and $\alpha_{12}$.

\subsection{Effective Potential Theory}
\label{sec:ept}
The effective potential theory (EPT, \citet{baalrud13, baalrud15}) extends the range of validity of the Boltzmann equation and the associated 
Chapman-Enskog solution for ionic transport coefficients to
strong ion coupling by rigorously including their correlations in the pair potential. This accounts for the presence of the surrounding ions in a collision
between two ions. This was recently derived from first principles \citep{baalrud19} 
with an expansion in terms of the departure of correlations from their equilibrium values rather than in terms of the strength of the correlations.
This gives a kinetic equation that is similar to the Boltzmann equation but in which the pair potential $V_{ij}(r)$ is
replaced by the potential of mean force, $V_{ij}^{\rm eff}(r)$. By definition, the potential of mean force is related to the pair distribution
function  
\begin{equation}
	g_{ij}(r) = \exp (-V^{\rm eff}_{ij}(r)/\kb T).
  \label{Vept}
\end{equation}
Given a pair interaction potential $V_{ij}(r)$, the pair distribution function 
can be extracted from classical MD simulations, or more economically with the integral theory of fluids (\citet{hmcd}, Chap. 4).
Inverting Equation (\ref{Vept}) gives $V^{\rm eff}_{ij}(r)$, which is then applied in the Chapman-Enskog collision integrals.  
In the limit of a weakly coupled system, $V^{\rm eff}_{ij}(r) \rightarrow V_{{\sss\rm D},ij}(r)$, the static screened Coulomb potential (Equation \ref{yukawa}), and the dilute gas 
limit is recovered exactly. 
The EPT has been extended to mixtures \citep{by14, daligault16, shaffer17}. Since the EPT can be solved for transport coefficients using the Chapman-Enskog formalism, 
the evaluation of the diffusion coefficient is very fast. It gives self-diffusion coefficients with an accuracy better than 9\%
for $\Gamma < 30$ for the one-component plasma \citep{baalrud15}, and matches ab initio simulations of a deuterium plasma to better than 6\% where
$\Gamma \le 9.5$ \citep{daligault16}. In general, the potential of mean force $V^{\rm eff}_{ij}(r)$ must be evaluated numerically even for a pair potential $V_{ij}(r)$
with a simple analytic form. The practical advantage of the single pre-tabulation of the Paquette et al. collision integrals for any binary mixture of ions 
is lost with the EPT.

The EPT theory is general and can be applied to systems interacting with any pair potential, as long as $g_{ij}(r)$ is known.  The most reliable  
$g_{ij}(r)$ come from ab initio simulations but then little is gained in terms of computational cost.  Instead, we use a recently developed model
for dense plasmas that combines an average atom model with the integral theory of fluids that provides both a pair potential and the
corresponding radial distribution function.

\subsubsection{Model for the ion-ion potential in dense, partially ionized plasmas}
\label{sec:aa-tcp}
In view of the computational cost of quantum MD simulations and the severe limitations of heuristic models of pressure ionization, we have adopted
a recently developed model of 
dense, partially ionized, partially degenerate, and weakly to strongly coupled plasmas that combines both physical realism and relative ease of computation. 
The average-atom, two-component plasma model (AA-TCP) considers a plasma composed of identical ions with an average charge ${\bar Z}$ with bound electrons in a sea of quantum mechanical
electrons of arbitrary degeneracy.  The model has no adjustable parameters and only requires the composition, temperature and density of the plasma
as inputs.  It provides a self-consistent solution for the energies and wave functions of the bound
and continuum states, the average ion charge, all correlation functions in the fluid, and the ion-ion pair potential. It
naturally accounts for strongly non-linear screening as well as pressure and temperature ionization and it can treat arbitrary mixtures without any additional 
approximation \citep{starrett13, starrett13err, starrett_is, starrett_mixtures}. 
The pair potential does not have a prescribed functional form such as the Yukawa potential but is calculated numerically within the model. 
The AA-TCP ion-ion potential $V_{ij}(r)$ can be used as an input to classical MD simulations, resulting in a dense plasma model called Pseudo-Atom Molecular 
Dynamics (PAMD, \cite{starrett_pamd}), or in the EPT model through the corresponding $g_{ij}(r)$ \citep{daligault16}. 
A limitation of this model is that it does not account for a distribution of ionic charge states.  All ions of a given species have the same average charge. 

The AA-TCP model is based on well-established theory but its formulation and numerical implementation are fairly elaborate. The model was developed and 
has evolved over several years. A recent review \citep{saumon20} provides a guide to the key publications. A more pedagogical introduction to the 
concepts and elements of such models is given in \cite{saumon14}. \cite{starrett_is} summarizes the final version of the model for a plasma with one ion species and
includes a discussion of several key numerical details.

The AA-TCP model has been validated by numerous comparisons with ab initio simulations and generally gives excellent results for the pair
distribution function -- a test of the quality of $V_{ij}(r)$ -- \citep{starrett13, starrett13err, starrett_is, starrett_mixtures}, the equation 
of state \citep{starrett_eos} and diffusion coefficients \citep{daligault16} 
over a wide range of temperatures and densities for elements ranging from hydrogen to tungsten, including 
binary mixtures.  It also compares very well with an accurate X-ray Thomson scattering experiment on warm dense aluminum \citep{starrett15_Al}. 
The combination of the AA-TCP model with the EPT thus opens the possibility of computing inter-diffusion coefficients
with a high degree of physical realism at a reasonable computational cost. While the combination of the EPT and AA-TCP models is more approximate than 
ab initio simulations in strongly coupled plasmas, it can handle systems with trace species and highly asymmetric mixtures without difficulty and is considerably more economical.

\begin{figure}[h]
	\epsscale{1.20}
   \plotone{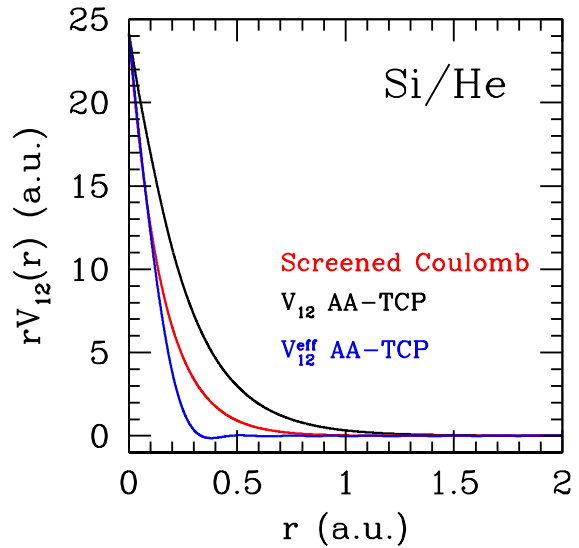}
   \caption{Comparison of Si-He pair potentials used in three different approaches to compute diffusion coefficients. The plasma is a mixture of He with a trace of Si ($x({\rm Si})=10^{-3}$)
	   at $\log T \, ({\rm K})=5.737$, $\log \rho \, ({\rm g/cm}^3) = 3.486$ (see Table \ref{tab:tab_cz_he}). 
	   The average ion charges, obtained with the AA-TCP model, are ${\bar Z}({\rm He})=Z_1=2.000$ and ${\bar Z}({\rm Si})=Z_2=12.121$ for all three potentials.
	   This plasma is strongly coupled ($\Gamma=15.259$) and strongly degenerate ($\kb T/\epsilon_{\sss\rm F}=0.014$).
	   For clarity, the product $rV_{12}(r)$ is shown so that as $r \rightarrow 0$, $rV_{12}(r)= Z_1 Z_2$. 
	   Under these conditions, the screened Coulomb potential (Equations \ref{yukawa}--\ref{lambda_sm16}) has reached the limit where the screening length 
	   $\lambda=a=0.152\,$ a.u. regardless of 
	   whether one uses the definition of \cite{paquette86a}, \cite{fbdt15} or \cite{stanton16}. Those three screened Coulomb potentials are thus identical here. $V_{12}$ is the ion-ion
   pair potential from the AA-TCP model  and $V_{12}^{\rm eff}$ is the corresponding potential of mean force (Equation \ref{Vept}). 
    [{\it See the electronic edition of the Journal for a color version of this figure.}]}
    \label{Vii}
\end{figure}

\subsubsection{Comparison of ion-ion potentials}

A comparison of the ion-ion potentials used in the above models illuminates how each accounts for the physics of correlations and screening in the plasma. For this purpose, we choose plasma conditions that 
emphasize the differences between the potentials and for which the diffusion coefficients vary significantly (see section \ref{sec:results_diff}). Figure \ref{Vii} shows 
potentials for a plasma of He (species 1)  with a trace of Si (species 2, $x_2=10^{-3}$) at the bottom of the convection zone of a low-$\teff$ white dwarf with a He envelope ($\log T \, ({\rm K}) = 5.737$ and
$\log \rho \, ({\rm g/cm}^3) = 3.486$). Under those conditions, the average ion charges obtained with the AA-TCP plasma model are ${\bar Z}({\rm He})=2.000$ and ${\bar Z}({\rm Si})= 12.121$. For the purpose of this
comparison, these values are applied to the calculation of all three potentials shown.  The coefficient of inter-diffusion of Si in a He-dominated plasma is determined by the Si-He pair potential 
while the plasma conditions (electron density and degeneracy) are dominated by the background He plasma.  The figure shows the product $rV_{12}(r)$ for clarity, giving a finite and 
common value of $Z_1 Z_2=24.242$ at $r=0$.  A pure Coulomb potential would appear as a horizontal line. The rapid decrease of $rV_{12}(r)$ away from the origin reflects the screening of 
the pure Coulomb interaction by electrons and ions. 
The Si-He pair potential $V_{12}(r)$ from the AA-TCP model in Figure \ref{Vii} accounts for electron screening only. In this case the screening is from the cloud of free electrons 
surrounding each ion (He and Si) obtained by solving the Schr\"odinger equation for free states including the interactions with other electrons and the surrounding ions.
This potential can describe the highly non-linear screening by free electrons in plasmas with strong electron-electron and electron-ion couplings, as well as the inherently non-linear bound states.
The AA-TCP pair potential is used in the classical molecular dynamics simulations (PAMD) that simulate the ion-ion correlations (i.e. ion screening) to all orders.  
Both electron and ion screening are included in the screened Coulomb potential (Equations \ref{yukawa} and \ref{debye}) but in the linear limit of weak coupling. For dense, strongly coupled
plasmas, the screening length is made to approach the ion sphere radius (Equations \ref{l_max}, \ref{lambda_fbdt15}, \ref{lambda_sm16}), which mimics the effect of strong ion screening
and corrects the failure of the weakly screened Coulomb potential.  This strong screening limit is reached in this particular example and all three choices of the screening length revert to 
the ion sphere radius and all three potentials are identical (given that they all use the same ion charges). 
In this approximate description, the introduction of ion correlations softens the pair potential significantly compared to the AA-TCP pair potential.
Finally, the effective potential $V_{12}^{\rm eff}(r)$ (Equation \ref{Vept}) is the  potential of mean force built from the AA-TCP pair potential 
$V_{12}(r)$ that accounts for electron screening and the pair distribution function that accounts for ion screening. This is the potential used in the EPT of transport coefficients and is the
least repulsive of all three. This comparison shows that the simple modifications of the screening length of \cite{paquette86a}, \cite{fbdt15}
and \citet{stanton16} do not properly account for strong ion screening except at very short range ($r\wig< 0.1\,$a.u.). In general a more repulsive potential results in a 
larger collisional cross-section and, as we will see below, smaller diffusion coefficients.  

\begin{deluxetable*}{lccc}[ht]
\tablecolumns{4}
\tablewidth{0pt}
\label{tab:transport_models}
\tablecaption{Comparison of the approximations in the models used to calculate diffusion coefficients herein}
\tablehead{
	\colhead{Quantity}        & \colhead{Paquette et al.}     & \colhead{Molecular Dynamics (PAMD)}                  &  \colhead{Effective Potential Theory}
}
\startdata
Ion charge                &  From a separate model    &  Implicit in input $V_{ij}(r)$  & Implicit in input $g_{ij}(r)$ \\
        ~                 &                           &  from AA-TCP model              &    from AA-TCP model \\[5pt]
Ion-ion pair potential    &  Static screened Coulomb   &  From AA-TCP model             &  Implicit in input $g_{ij}(r)$\\
			  &   ``Yukawa''               &                                      &           from AA-TCP model \\[5pt]
Order of the collisions   &  2-body                    &   N-body                             & 2-body and higher order \\[5pt]
Diffusion coefficients    &  Collision integrals       &  From particle                       & Collision integrals with \\
~                         &  with Yukawa $V_{ij}(r)$   &  trajectories                        &  effective $V_{ij}^{\rm eff}(r)$   \\[5pt]
Thermodynamic factor      &     $J=1$                  & From AA-TCP model                    &  Non-ideal               \\
                          &           ~                &         ~                            & (within EPT approximation) \\
\enddata
\end{deluxetable*}

\subsubsection{Summary of methods}
To summarize, there is a hierarchy of models and approximations that allow the computation of diffusion coefficients in white dwarf atmospheres.
In order of increasing physical sophistication and also of computational cost, the models discussed here are 1) the Chapman-Enskog theory of transport in dilute gases 
with collision integrals for
Coulomb potentials with an appropriate radial cutoff, 2) the \citet{paquette86a} approach that uses the same formalism but with a
static screened Coulomb potential, 3) the EPT theory that extends the Chapman-Enskog formalism to strongly coupled plasmas
by including ion correlations in the pair potential, 4) classical molecular dynamics, and 5) ab initio molecular dynamics.
The EPT can be applied to any pair potential such as pure Coulomb, Yukawa and more sophisticated potentials.  The first three
approaches can all be used from the very weak coupling limit up to various degrees of coupling, and have low
computational costs. Classical MD can be used
with any repulsive potential and has no physical approximation other than those implicit in the potential but is much more costly. It works well in the moderate 
to very strong coupling regime. We use classical MD below to validate the EPT results. The ab initio methods are based on classical MD for the ions 
but rather than
using a prescribed ion-ion potential, they use a model for calculating the structure of the electron fluid in the three-dimensional simulation box and the resulting forces on the ions.
They are thus much more expensive. The simplest form of ab initio simulation that treats the electrons explicitly is the Thomas-Fermi orbital free MD 
which is suitable for hot and dense plasmas.  Finally,
the most accurate and most expensive method is quantum MD, where the electrons are modeled quantum mechanically. For computational reasons
it is limited to rather low temperatures, typically $\lesssim 10^5\,$K. The main features of the three models we apply here are summarized in Table \ref{tab:transport_models}.

Computationally, the evaluation of the diffusion coefficients of \cite{paquette86a} and \cite{stanton16} are very fast as they require only 
the evaluation of a fit to pre-evaluated collision integrals for the analytic screened Coulomb potential. The EPT 
is slower because it evaluates the collision integrals using the potential of mean force which is not pre-determined but this is relatively fast, given a potential. In both cases, however, 
a model of the equation of state must be run to obtain the ion charges and/or the potential, which is far more costly than the evaluation of the collision integrals. The AA-TCP model 
is a reliable and advanced model for dense, partially ionized plasmas but its evaluation can take a few minutes to an hour per $(\rho,T)$ point, depending on the conditions 
and the atomic number of the elements in the mixture. For practical applications, diffusion coefficients evaluated with the combination of the AA-TCP model and of the EPT must be
pre-tabulated for the specific mixture. Finally, ab initio methods typically take 2 orders of magnitude longer than the AA-TCP model for evaluating the equation of state/potential and even longer to
obtain diffusion coefficients.

\section{Coefficients of inter-diffusion and ionic thermal diffusion}
\label{sec:results_diff}
Our main purpose is to revisit the calculation of diffusion coefficients under the conditions found in white dwarf envelopes with the EPT using AA-TCP 
potentials and compare with the values from the more approximate model of \citet{paquette86a}.  We also present a validation of the EPT results  against
classical MD results that use the same ion-ion potential $V_{ij}(r)$ that gives $V_{ij}^{\rm eff}(r)$ for the EPT calculation. This tests the accuracy of the EPT but not that of the input
potential that must be validated separately.

Cool white dwarfs develop a surface convection zone where the mixing time scale is much shorter than the diffusion time, resulting in a homogeneous composition throughout the
convective region. In the simplest picture, the abundances of heavy elements in the convection zone (which are observable in the star's spectrum) decrease as those ions  diffuse below the bottom of 
the convection zone. Thus we focus on diffusion at the bottom of the convection zone as the most relevant regime for the spectral evolution of white dwarf stars. 

We calculate the coefficient of inter-diffusion $D_{12}$ and the ionic contribution $\alpha_{12}$ to the thermal diffusion factor $\alpha_{\sss T}$ (Equation \ref{w12}) in white dwarf models with both pure 
hydrogen and pure helium envelopes. We consider the diffusion of traces of silicon and calcium which are two well-observed elements in the 
spectra of metal-polluted WDs \citep{zeidler86, dupuis93, dufour07}. The properties of the convection zone are taken from selected models along two white dwarf cooling
sequences with a mass of $M_\star=0.6\,$M$_\odot$, a pure carbon core, a He layer with a mass fraction of $10^{-2}\,M_\star$ and, for the H case, a superficial 
hydrogen layer with a mass of $10^{-4}\,M_\star$. The convection is modeled with the ML2 parametrization of the mixing length theory. 
These evolution models are described in \cite{fbb01}. \footnote{In particular, the DA sequence corresponds to that available at \tt{www.astro.umontreal.ca/$\sim$bergeron/CoolingModels/} } 
For each model in a sequence, the 
effective temperature, radius, gravity, temperature and density at the bottom of the convection zone, and fractional mass of the convection zone are given in Tables 
\ref{tab:tab_cz_h} 
(H case) and \ref{tab:tab_cz_he} (He case). The last two columns give the values of two important plasma parameters, the electron degeneracy parameter 
$\kb T/\epsilon_{\sss F}$, where $\epsilon_{\sss F}$ is the Fermi energy, and the plasma coupling parameter $\Gamma$ (Equation \ref{gamma}).
Both parameters are evaluated for a plasma composed of the dominant background plasma species (H or He). At the bottom of the hydrogen convection zone, the plasma is 
weakly to partially degenerate ($\kb T/\epsilon_{\sss F} \sim 6$ -- 0.4 and moderately coupled ($\Gamma \sim 0.3$--0.6).  In the coolest helium envelopes, however, the plasma at the bottom of the convection zone can
become strongly degenerate ($\kb T /\epsilon_{\sss F} \ll 1$) and strongly coupled ($\Gamma \gtrsim 10$).

For each set of conditions listed, we run the AA-TCP model for dense, partially ionized plasmas with a trace abundance of the heavy element ($x_2=10^{-3}$) in
a plasma of H or He. This provides self-consistent average ion charges ${\bar Z_i}$, pair potentials $V_{ij}(r)$, pair distribution functions $g_{ij}(r)$, and structure factors 
$S_{ij}(k)$. These quantities are applied to the calculations of $D_{12}$ and $\alpha_{12}$ described below. 

\begin{deluxetable*}{cccccccc}
\label{tab:tab_cz_h}
\tablecolumns{8}
\tablewidth{0pt}
\tablecaption{Physical conditions at the bottom of the H convection zone of a $M_\star=0.6\,$M$_\odot$ DA white dwarf (see text for details).}
\tablehead{
	\colhead{ $\teff$ (K) } &  \colhead{ $\log R{\sss\rm CZ}\,$(cm)}  &  \colhead{ $\log g_{\sss CZ}$(cm/s$^2$) } &  \colhead{ $\log T$ (K) }   &  \colhead{ $\log \rho$ (g/cm$^3$) }  &  \colhead{ $\log M_{\sss CZ}/M_\star$ }  &  \colhead{ $\kb T/\epsilon_{\sss F}$ }   &  \colhead{ $\Gamma$ }
}
\startdata
 8499       & 8.9433 & 8.0166 &  5.651  & -0.883  &  -9.151  &     5.772  &    0.257  \\
 7989       & 8.9422 & 8.0190 &  5.696  & -0.645  &  -8.874  &     4.444  &    0.279  \\
 7509       & 8.9411 & 8.0213 &  5.732  & -0.446  &  -8.641  &     3.557  &    0.299  \\
 7060       & 8.9401 & 8.0235 &  5.763  & -0.273  &  -8.437  &     2.931  &    0.317  \\
 6522       & 8.9389 & 8.0262 &  5.806  & -0.042  &  -8.163  &     2.272  &    0.343  \\
 6282       & 8.9382 & 8.0277 &  5.837  &  0.101  &  -7.989  &     1.956  &    0.357  \\
 6064       & 8.9375 & 8.0294 &  5.880  &  0.286  &  -7.761  &     1.626  &    0.372  \\
 5866       & 8.9368 & 8.0313 &  5.928  &  0.486  &  -7.511  &     1.336  &    0.389  \\
 5685       & 8.9360 & 8.0336 &  5.986  &  0.714  &  -7.223  &     1.075  &    0.405  \\
 5517       & 8.9351 & 8.0363 &  6.047  &  0.963  &  -6.907  &     0.845  &    0.426  \\
 5354       & 8.9341 & 8.0390 &  6.095  &  1.188  &  -6.624  &     0.668  &    0.454  \\
 5208       & 8.9331 & 8.0418 &  6.133  &  1.398  &  -6.361  &     0.528  &    0.489  \\
 5118       & 8.9325 & 8.0435 &  6.145  &  1.514  &  -6.222  &     0.454  &    0.520  \\
 4993       & 8.9316 & 8.0454 &  6.131  &  1.633  &  -6.096  &     0.366  &    0.588  \\
\enddata
\end{deluxetable*}

\begin{deluxetable*}{cccccccc}
\label{tab:tab_cz_he}
\tablecolumns{8}
\tablewidth{0pt}
\tablecaption{Physical conditions at the bottom of the He convection zone of a $M_\star=0.6\,$M$_\odot$ DB white dwarf (see text for details).}
\tablehead{
	\colhead{ $\teff$ (K) } &  \colhead{ $\log R_{\sss\rm CZ}$(cm) }  &  \colhead{ $\log g_{\sss CZ}$(cm/s$^2$) } &  \colhead{ $\log T$ (K) }   &  \colhead{ $\log \rho$ (g/cm$^3$) }  &  \colhead{ $\log M_{\sss CZ}/M$ }  &  \colhead{ $\kb T/\epsilon_{\sss F}$ }   &  \colhead{ $\Gamma$ }
}
\startdata
20382       & 8.9487 & 8.0053 &  5.961  & -0.478  &  -8.835  &     9.998  &    0.434  \\
18025       & 8.9433 & 8.0186 &  6.349  &  1.160  &  -6.817  &     1.978  &    0.625  \\
17703       & 8.9426 & 8.0203 &  6.380  &  1.290  &  -6.656  &     1.739  &    0.643  \\
16720       & 8.9406 & 8.0252 &  6.456  &  1.612  &  -6.258  &     1.262  &    0.692  \\
15085       & 8.9375 & 8.0324 &  6.530  &  1.967  &  -5.824  &     0.869  &    0.765  \\
12471       & 8.9329 & 8.0430 &  6.581  &  2.365  &  -5.350  &     0.530  &    0.923  \\
10039       & 8.9287 & 8.0518 &  6.523  &  2.664  &  -5.042  &     0.293  &    1.327  \\
 8784       & 8.9268 & 8.0546 &  6.372  &  2.734  &  -5.032  &     0.186  &    1.983  \\
 7517       & 8.9249 & 8.0582 &  6.146  &  2.859  &  -4.916  &     0.091  &    3.675  \\
 6603       & 8.9231 & 8.0637 &  5.945  &  3.071  &  -4.604  &     0.041  &    6.865  \\
 5612       & 8.9192 & 8.0798 &  5.737  &  3.486  &  -3.914  &     0.014  &   15.259  \\
\enddata
\end{deluxetable*}

We perform calculations with the Chapman-Enskog collision integrals formalism\footnote{This calculation and the EPT calculation use the second order approximation 
to $D_{12}$ and the first approximation to $\alpha_{12}$ of the Chapman-Enskog theory.} with a static screened Coulomb potential as described in \citet{paquette86a}
and updated in \citet{fbdt15} with the screening length given by Equations \ref{lambda_fbdt15} and \ref{TF_length}.
The ion charges are obtained from an EOS based on the occupation probability formalism \citep{hm88}.
Although it uses a different choice of screening length and ionization model than \citet{paquette86a}, this updated model is 
hereafter referred to as ``Paquette et al.''.  To illustrate the importance of 
the ionization model in evaluating diffusion coefficients, we also evaluate the Paquette et al. coefficients using the ion 
charges obtained with the AA-TCP model (section \ref{sec:aa-tcp}). Due to the factorization of the static screened Coulomb potential in this formalism, 
these two calculations use the same pre-tabulated dimensionless collision integrals and differ only in the choice of ion charges.

Given the ion-ion pair potentials $V_{ij}(r)$ from the AA-TCP model, we evaluate the coefficient of inter-diffusion from classical molecular dynamics 
simulations \citep{starrett_pamd, daligault16}. 
The thermodynamic factor $J$ is calculated from the structure factors $S_{ij}(k)$  obtained from the AA-TCP model (Equation \ref{thermo_factor_2}). For trace species at 
the conditions at the bottom 
of the convection zone (Tables \ref{tab:tab_cz_h} and \ref{tab:tab_cz_he}), this correction remains small with $1 \le  J < 1.02$.  The molecular dynamics simulations were performed in 
a cubic box containing $N=50\,000$ particles (with $N_2=250$) with periodic boundary conditions. 
The time step was set to $\Delta t=0.005 \omega_p^{-1}$. 
The plasma frequency of the light, background ions $\omega_p=\sqrt{4\pi n_iZ^2 e^2 / m_i}$ 
is a measure of the shortest characteristic dynamic time scale of the plasma.  By choosing such a short time step, the motions of both the light and 
heavier ions in the mixture are very well resolved. After an equilibration period of $10^5$ time steps, the  trajectories of the ions were followed for an additional $8.4 \times 10^6$ time steps,
which were used to calculate $D_{12}$ from Equation \ref{diff_micro}.  
In these simulations, the abundance of the trace heavy element (Ca or Si) is set to $x_2 = 0.005$ or 0.01 depending on the $(T, \rho)$ conditions.  
These are the largest values of $x_2$ that give converged values of $D_{12}$ while staying within the limit of a trace abundance. After a long time, the integral in Equation \ref{diff_micro}
approaches a constant value. The value of $D_{12}$ is obtained by averaging the running integral over a block of time after it has reached a plateau.
The statistical uncertainty on $D_{12}$ is estimated from the dispersion around this average value.

Finally, we evaluate $D_{12}$ and the $\alpha_{12}$ with the Effective Potential Theory \citep{baalrud13, baalrud15} applied to the collision integrals, 
with the pair distribution functions $g_{ij}(r)$ from the AA-TCP model as input. 

\begin{figure}[h]
	\epsscale{1.20}
   \plotone{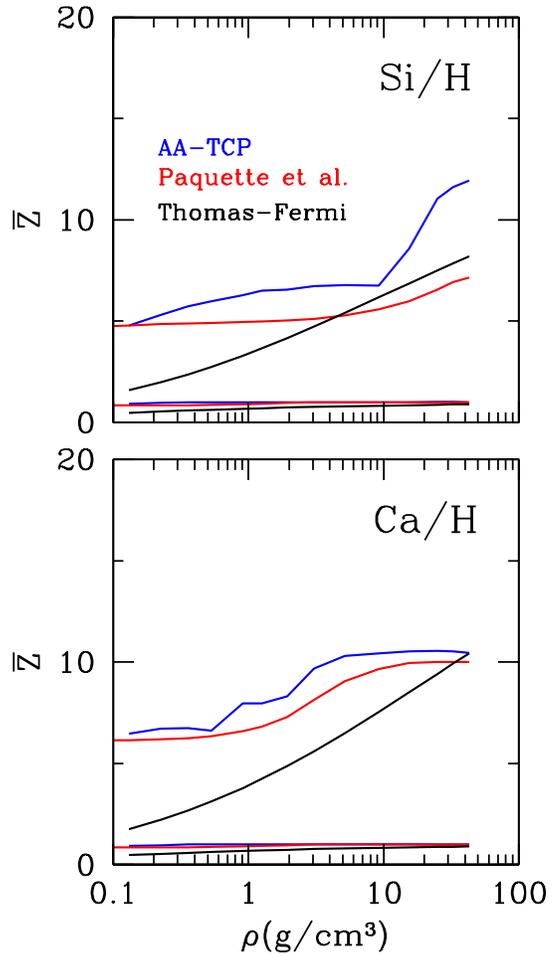}
   \caption{Average ion charges in a plasma of a trace ($x_2=0.001$) of silicon (upper panel) and calcium (lower panel) in hydrogen. The $(\rho, T)$ conditions
	    correspond to the bottom of the superficial hydrogen convection zone of a 0.6$\,$M$_\odot$
	    DA white dwarf cooling from $\teff= 8500\,$K (left) to $5000\,$K.  The abscissa is the density at the bottom
	    of the convection zone, which increases as the WD cools (Table \ref{tab:tab_cz_h}).  
	    The lower set of curves in each panel shows 
	    the charge of hydrogen, the upper set that of the heavier ion (Si or Ca).  The charges computed from three models are shown: with the
	    model used to calculate the Paquette et al. coefficients \citep{fbdt15} (red), the AA-TCP plasma model \citet{starrett_is} (blue),
	    and the simple Thomas-Fermi model (black, \cite{stanton16}). 
	    The first two models predict full ionization of H ($\bar{Z}=1$) but differ in their prediction of the charge of the heavier ion, most notably for Si.
            The lack of electronic shell structure in the Thomas-Fermi model results in a featureless increase of $\bar Z$.
    [{\it See the electronic edition of the Journal for a color version of this figure.}]}
    \label{Zbar_in_H}
\end{figure}

\begin{figure}[h]
   \epsscale{1.20}
   \plotone{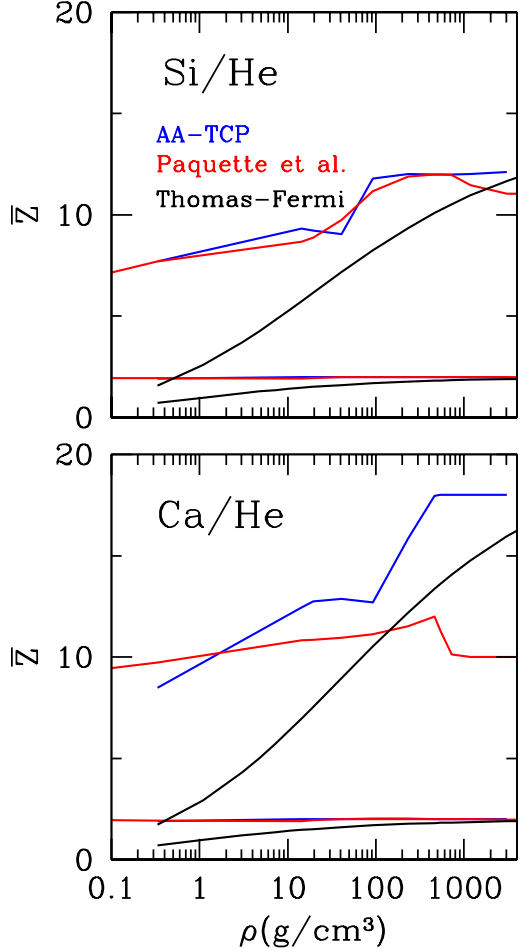}
   \caption{Same as Figure \ref{Zbar_in_H} but for traces of Si and Ca at the bottom  of the superficial helium convection zone (Table \ref{tab:tab_cz_he}).
   Both the Paquette et al. and the AA-TCP models predict that helium is fully ionized under these conditions ($\bar Z =2$). At the highest 
	   densities, Si retains only its $1s^2$ electrons ($\bar Z \sim 12$)  in both models. For Ca, the AA-TCP model also predicts that only the $1s^2$ electrons 
	   remain bound.  On the other hand, the model used by Paquette et al. predicts
	   a much lower charge of ($\bar{Z} \sim 10$) for the Ca ion, retaining the full $n=2$ electronic shell.       
           The lack of electronic shell structure in the Thomas-Fermi model results in a featureless increase of $\bar Z$.
           [{\it See the electronic edition of the Journal for a color version of this figure.}]}
    \label{Zbar_in_He}
\end{figure}

The charges of the Ca and Si ions at the bottom of the hydrogen convection zone are shown in Fig. \ref{Zbar_in_H}. In this and subsequent figures, the density is used as 
the independent variable following the bottom of the convection zone for a cooling sequence of white dwarfs. The temperature also varies along the ordinate (see Tables
\ref{tab:tab_cz_h} and \ref{tab:tab_cz_he}). Under these conditions, hydrogen is always essentially fully ionized,  
the AA-TCP model predicting a slightly higher degree of ionization (by $\Delta {\bar Z} \lesssim 0.1$) at lower densities ($\sim 0.1$ - 1\,g/cm$^3$).  The degree of ionization of Si 
in H is systematically larger in the AA-TCP model, especially at the higher density/temperatures where only the $1s^2$ electrons remain bound (${\bar Z}{\rm(Si)} \sim 12$) while the 
\citet{fbdt15} model predicts that it also retains most of the L-shell electrons (${\bar Z}$(Si)$\sim 7$).  For Ca, the pattern of ionization is similar, with an increase 
toward higher densities and temperatures, but in this case both models predict nearly the same average charge for the Ca ions with the AA-TCP being systematically higher.
In helium envelopes, the convection zone can reach to densities well above $10^3\,$g/cm$^3$ where the plasma is strongly coupled and strongly degenerate. In all these cases, 
both models predict that He is fully ionized (Fig. \ref{Zbar_in_He}). Interestingly, they
predict very similar degrees of ionization for Si over this wide range of conditions while the Ca charges differ markedly above 100$\,$g/cm$^3$, 
with the AA-TCP model again predicting that Ca retains only its $1s^2$ electrons (${\bar Z}$(Ca)=18) and the simpler ionization model reaching a closed-shell 
Ne-like  configuration (${\bar Z}$(Ca)=10).
The simplest model to calculate ion charges while taking into account temperature and pressure ionization is the semi-classical Thomas-Fermi average atom model 
(\cite{fmt49}, see \cite{stanton16} for a practical calculation of the ion charge). This model fares poorly at low density and temperatures where it predicts that H and He 
are only about 50\% ionized. For heavier elements, the absence of electronic shell structure in the Thomas-Fermi model provides a very smooth transition towards full ionization
but predicts ion charges that are quite different from the other two models over nearly the full range of conditions shown in Figures \ref{Zbar_in_H} and \ref{Zbar_in_He}. 
Note that any reasonable ionization model must reach full ionization at high temperatures and high densities, hence the influence of the choice of 
ionization model on the diffusion coefficient must vanish at large enough depth in the star. 

\begin{figure}[h]
   \epsscale{1.20}
   \plotone{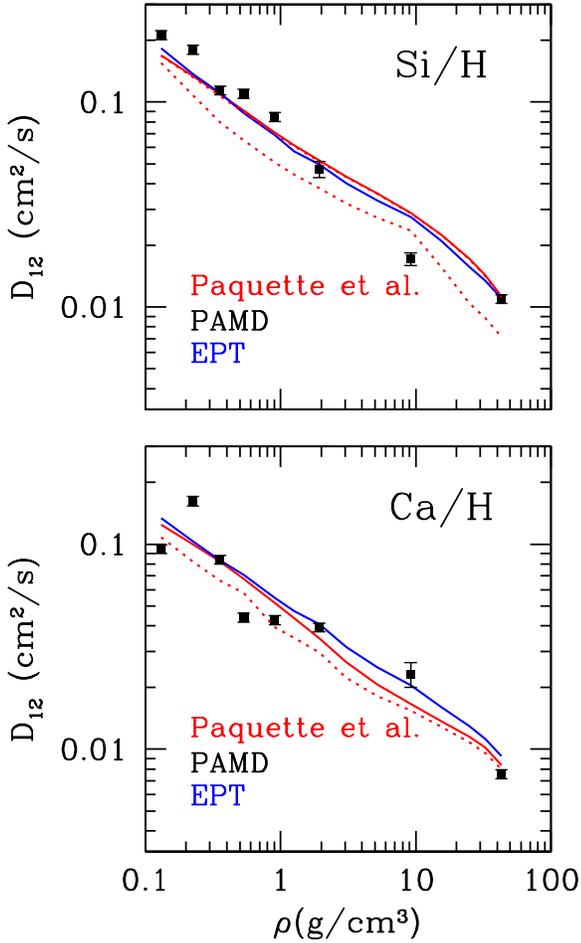}
   \caption{Coefficient of inter-diffusion of Si and Ca at the bottom of the superficial hydrogen convection zone (Table \ref{tab:tab_cz_h}).  
	    Four calculations are shown: The value from 
	    Paquette et al. (red solid lines), from the PAMD classical molecular dynamics simulations (black squares), and from the
	    effective potential theory (EPT, blue solid curves). The latter two calculations are based on the same ion-ion pair potential and
	    ideally should give the same results. The Paquette et al. diffusion coefficient calculated with the ionic charges from the
	    AA-TCP model are shown by the red dotted line. Under these conditions, the plasma is weakly coupled and the EPT agrees
	    very well with Paquette et al.
    [{\it See the electronic edition of the Journal for a color version of this figure.}]}
    \label{diff_in_H}
\end{figure}

Figure \ref{diff_in_H} shows the coefficient of inter-diffusion of Si and Ca in hydrogen envelopes. The general trend of $D_{12}$ at the bottom of the convection zone of stars with decreasing $\teff$
is of a rapid decrease caused primarily by the increased density and coupling of the plasma. Larger densities and plasma coupling implies more frequent and  stronger collisions, respectively,
that both inhibit diffusion.  Four calculations of $D_{12}$ are shown for both Si and Ca. For Si, the 
Paquette et al. result (solid red)  and the EPT results (solid blue) agree remarkably well over the full range of conditions. This agreement is somewhat fortuitous, however. If the ion charges
from the AA-TCP model that  are used in the EPT calculation are also applied in the Paquette et al. model, $D_{12}$ decreases by $\sim 25$\%. Thus for a given $Z$(Si), the static screened Coulomb potential of 
Paquette et al. leads to an underestimate of $D_{12}$ by about 25\%.  The classical molecular dynamics simulations (black squares), which are based on 
the same AA-TCP interaction potentials $V_{ij}(r)$ that leads to the potential of mean force $V_{ij}^{\rm eff}(r)$ used in the EPT calculation show a level of scatter 
that is greater than their formal statistical uncertainties (shown by error bars). This illustrates the difficulty in estimating accurate diffusion coefficients 
of trace species from molecular dynamics simulations.
Nonetheless, the latter agree quite well with the EPT calculation within the scatter.   The diffusion coefficient of Ca (lower panel of Fig. \ref{diff_in_H}) shows the same general features.
A comparison with $D_{12}$ evaluated with the fits of \cite{stanton16}, using the same ion charges as the Paquette et al. calculation (solid red curve) shows remarkable agreement with differences
that are almost always under 2\% and no more than 3\%, for both Ca and Si. 

\begin{figure}
   \epsscale{1.20}
   \plotone{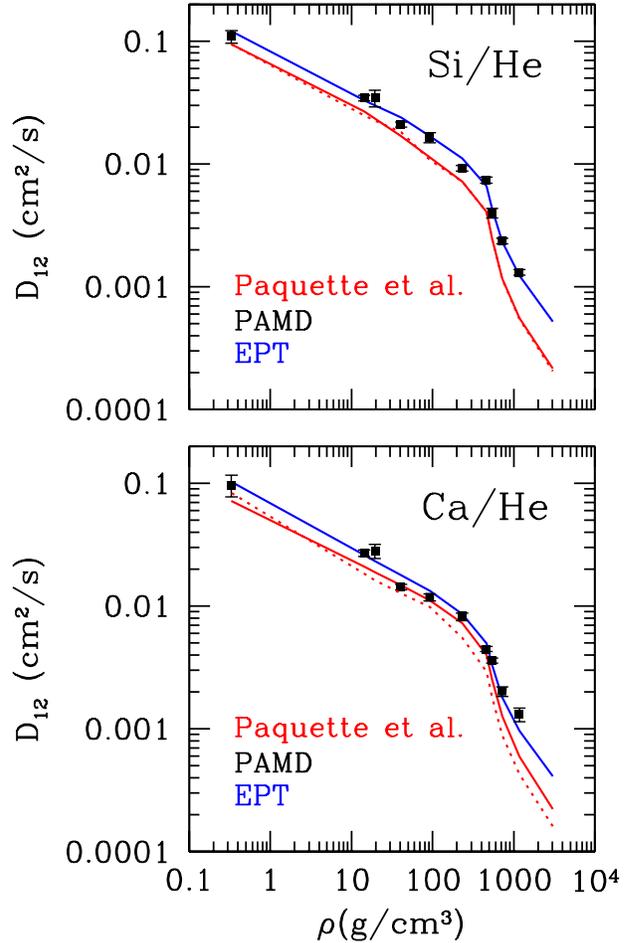}
   \caption{Same as Figure \ref{diff_in_H} but for the diffusion of Si and Ca at the bottom of the superficial hydrogen convection zone (Table \ref{tab:tab_cz_he}).  
	   At the higher densities encountered in the cooler He WDs, the plasma is more strongly coupled and several assumptions in  \cite{paquette86a} are
	    no longer valid, resulting in an underestimation of the diffusion coefficient by up to a factor of 2.4.
	    Note the different scale of the ordinate axis compared to Fig \ref{diff_in_H}.
    [{\it See the electronic edition of the Journal for a color version of this figure.}]}
    \label{diff_in_He}
\end{figure}

\begin{figure}
   \epsscale{1.20}
   \plotone{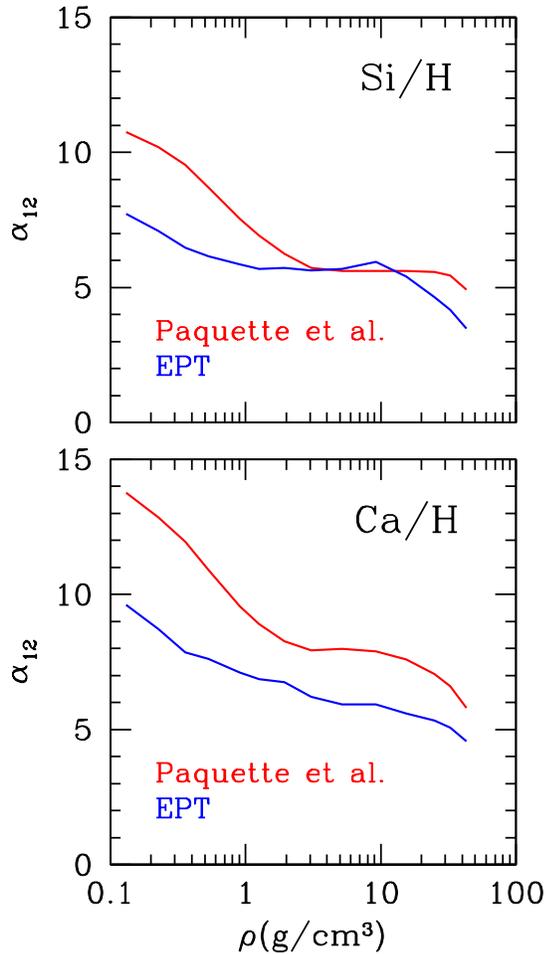}
   \caption{Coefficient of ionic thermal diffusion factor of Si and Ca at the bottom of the superficial hydrogen 
	   convection zone (Table \ref{tab:tab_cz_h}) from the Paquette et al. and the EPT models.
            [{\it See the electronic edition of the Journal for a color version of this figure.}]}
    \label{a12_in_H}
\end{figure}

\begin{figure}
   \epsscale{1.20}
   \plotone{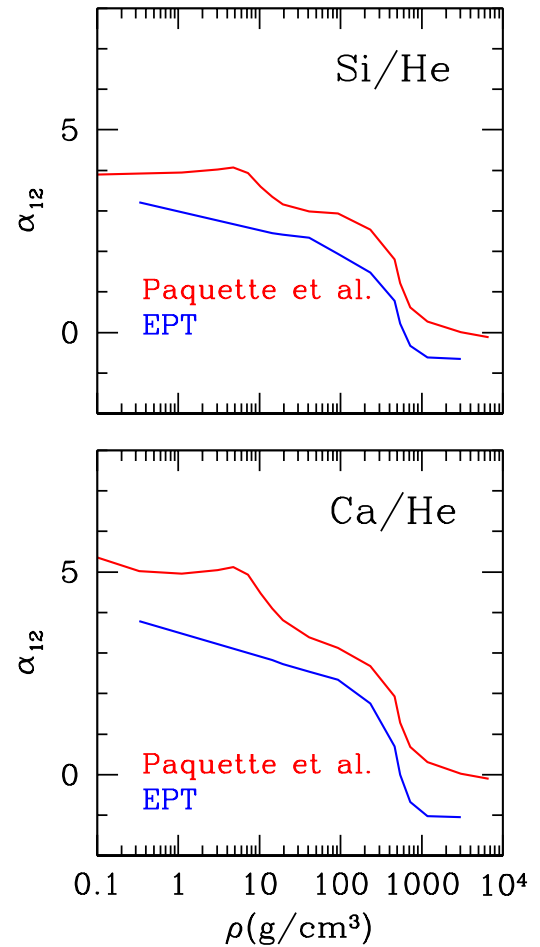}
   \caption{Same as Fig. \ref{a12_in_H} but showing the coefficient of ionic thermal diffusion factor of Si and Ca at the bottom 
	   of the superficial helium convection zone (Table \ref{tab:tab_cz_he}).
            [{\it See the electronic edition of the Journal for a color version of this figure.}]}
    \label{a12_in_He}
\end{figure}

The coefficients of inter-diffusion of Si and Ca in helium envelopes behave in a similar fashion (Fig. \ref{diff_in_He}) but  differ from that in hydrogen envelopes. 
At densities above 400$\,$g/cm$^3$, $D_{12}$ drops rapidly due to the rapid increase of the Coulomb coupling  between the heavy ions and the 
He$^{2+}$ plasma  as the convection zone in the He envelope reaches much deeper into the star.
The scatter in the simulations is reduced compared to the hydrogen case because of the 
stronger coupling, which results in faster equilibration and convergence as well as smaller fluctuations.  The EPT calculation is validated by the excellent agreement with the 
classical molecular dynamics simulations.  As in the case of hydrogen, the Paquette et al. model systematically underestimates $D_{12}$ (more for Si than for Ca) by $\sim 35$\% and
up to a factor of 2.4 in  He-envelope WDs with $\teff \sim 5600\,$K. This can be traced back to the more repulsive screened Coulomb potential compared to the potential 
of mean force (Figure \ref{Vii}). Interestingly, in this case the inter-diffusion coefficients of \cite{stanton16} agrees with Paquette et al. 
(solid red curves) to better than 1\% at densities above 400$\,$g/cm$^3$ but they deviate from each other by $\sim 15$\% at the lowest density of 0.3$\,$g/cm$^3$.

Figures \ref{a12_in_H} and \ref{a12_in_He} show the ionic contribution $\alpha_{12}$ to the thermal diffusion factor $\alpha_{\sss T}$. Because it is defined as a pre-factor to $D_{12}$,
most of the $\rho$--$T$ dependence of the ionic thermal diffusivity is taken up by $D_{12}$ and $\alpha_{12}$ depends only weakly on the plasma conditions and the charge of the heavy ion. 
It decreases steadily as the plasma coupling increases.  For hydrogen envelopes, $\alpha_{12}$ is 2--3 times larger than in helium envelopes. We find that the Paquette et al.  and EPT models 
are in generally good 
agreement although the former systematically overestimates $\alpha_{12}$ by 20--40\%. In particular, the EPT model predicts that $\alpha_{12}$ changes sign 
in the strongest coupling regimes encountered in the He envelopes (Figure \ref{a12_in_He}). This also happens in the \cite{paquette86a}  calculation but to a much smaller extent.
A negative $\alpha_{12}$ implies that thermal diffusion will tend to make the heavy species move upward in the star. However, the ionic thermal diffusion term is
typically 10--30 times smaller than the gravitational settling terms in Equation \ref{w12} under the conditions of interest.

These calculations show that {\it for the same ionic charges}, the AA-TCP plasma model combined with the EPT gives diffusion coefficients $D_{12}$ that are systematically 
higher than those of \cite{paquette86a}. The
difference is modest $\lesssim 35$\% when the plasma coupling is moderate as in all H envelopes but grows to a factor of $\gtrsim 2$ in cool He envelopes ($\teff \lesssim 15000\,$K) where the 
plasma is strongly coupled. The dimensionless ionic thermal 
diffusion factor  $\alpha_{12}$ is generally lower in the EPT calculation by an amount that is essentially independent of the plasma coupling.  The dependence of $D_{12}$ on the charge of the heavy ion
(the light ions H or He being fully ionized at the bottom of the convection zone) is significant. The ionization model of \cite{fbdt15} used here to compute the nominal \cite{paquette86a} 
diffusion coefficients gives charges that can be in very good agreement with those of the AA-TCP plasma model but are typically lower, and sometimes considerably so. This tends 
to compensate for the intrinsic difference in the theory used to evaluate $D_{12}$. This is a cautionary statement about the importance of the ionization model and the
underlying equation of state model in evaluating transport coefficients in white dwarf envelopes.

\section{Diffusion time scales at the bottom of the convection zone}
\label{sec:results_time}

One of the most remarkable recent developments in the field of white dwarfs is the recognition that the presence of metal lines in the spectra of DZ, DBZ and DAZ white dwarfs 
originates from the accretion of planetary material  \citep{jura03, jura14, farihi16}. 
This provides a unique window into the detailed elemental composition of the accreted planetary solids that is not otherwise accessible from the observation of exoplanets.
The observed photospheric abundance pattern of metals results from the interplay of the composition of the accreted material, the accretion rate, and the processes of convective mixing and diffusion.
Thus, the composition of the infalling material that is deduced is sensitive to the relative diffusion time scale of the various accreted elements. As an illustration, we consider the evolution of the
surface mass fraction $X_2$ of a trace heavy element in the convection zone of a white dwarf that is accreting at a constant rate $\dot{M_2}$ \citep{dupuis93} 
\begin{equation}
	\frac{dM_2}{dt} = \frac{d(X_2 M_{\sss\rm CZ})}{dt} = \dot{M_2} + 4\pi R_{\sss\rm CZ}^2 \rho X_2 w_{12}.
\end{equation}
If we assume that $M_{\sss\rm CZ}$, $R_{\sss\rm CZ}$ and $w_{12}$ are constant, the solution is
\begin{equation}
	X_2(t) = X_2(0)e^{-t/\tau_{\rm d}} + \frac{\tau_{\rm d} \dot{M_2}}{M_{\sss\rm CZ}}(1 - e^{-t/\tau_{\rm d}})
\end{equation}
where $\tau_{\rm d}$ is the diffusion time scale of species 2 at the bottom of the convection zone, given by
\begin{equation}
	\tau_{\rm d}=\frac{-M_{\sss\rm CZ}}{4\pi R_{\sss\rm CZ}^2} \frac{1}{ \rho w_{12}}.  
\end{equation}
This diffusion time scale can be evaluated from the quantities given in Tables \ref{tab:tab_cz_h} and \ref{tab:tab_cz_he} and $w_{12}$ using Equation \ref{w12}.  Note that for downward 
diffusion, $w_{12} < 0$ and $\tau_{\rm d} > 0$.  At late times ($t \gg \tau_{\rm d}$) the abundance of the trace metal reaches an equilibrium value of 
\begin{equation}
	X_2^{\rm eq} = \frac{\tau_{\rm d} \dot{M_2}}{M_{\sss\rm CZ}}.
	\label{x2_equil}
\end{equation}
After accretion has stopped ($\dot{M_2}=0$) the mass fraction decreases exponentially
\begin{equation}
	X_2(t) = X_0 e^{-t/\tau_{\rm d}}.
\end{equation}
\cite{dupuis93} and \cite{bauer18}  have studied the evolution of such a trace metal abundance in a white dwarf undergoing episodic accretion.

We are interested in how our improved model of diffusion in dense plasmas affects the {\it relative} diffusion time scale of Ca and Si and the inferred 
composition of the accreted planetary material. 
In the accretion/diffusion equilibrium limit, the observed ratio of mass fractions of the accreted material is related to the ratio of diffusion time scales
\begin{equation}
\frac{X_{\sss\rm Si}}{X_{\sss\rm Ca}} = \frac{\tau_{\rm d}({\rm Si})}{\tau_{\rm d}({\rm Ca})} \frac{\dot{M}_{\rm Si}}{\dot{M}_{\rm Ca}},
\end{equation}
and in the case of no accretion, this ratio evolves exponentially on a time scale of the order of the shorter of the two diffusion time scales
\begin{equation}
	\frac{X_{\sss\rm Si}}{X_{\sss\rm Ca}} = \frac{X_0({\rm Si})}{X_0({\rm Ca})} \exp \bigg[- t \Big(\frac{1}{\tau_{\rm d}({\rm Si})} - \frac{1}{\tau_{\rm d}({\rm Ca})} \Big) \bigg].
\end{equation}

For the present purpose, we approximate the diffusion velocity by neglecting 
the ordinary and thermal diffusion terms\footnote{The ordinary diffusion term is negligible for a trace species \citep{fm79} and the thermal diffusion term is small \citep{paquette86b}. The ratio of the
diffusion time scales is even less sensitive to these approximations than $w_{12}$.} 
\begin{equation}
	w_{12} = D_{12} \Big[ \Big( \frac{Z_2}{Z_1}A_1 - A_2 \Big) \frac{m_0 g_{\sss\rm CZ}}{\kb T}  + \Big( \frac{Z_2}{Z_1} - 1 \Big) \frac{\partial \ln P_i}{\partial r} \Big].
	\label{w12_approx}
\end{equation}
For a given stellar structure, the ratio $\tau_{\rm d}$(Si)/$\tau_{\rm d}$(Ca) depends only on the values of $D_{12}$ and $Z_2$ for each element. The charge $Z_2$ also enters indirectly through $D_{12}$ 
(Figures \ref{diff_in_H} and \ref{diff_in_He}).  As we have seen above, a larger ion charge for the trace element $Z_2$ results in a smaller  $D_{12}$, a smaller 
pre-factor in Equation \ref{w12_approx}, and a longer diffusion time scale.

\begin{figure}
   \epsscale{1.20}
   \plotone{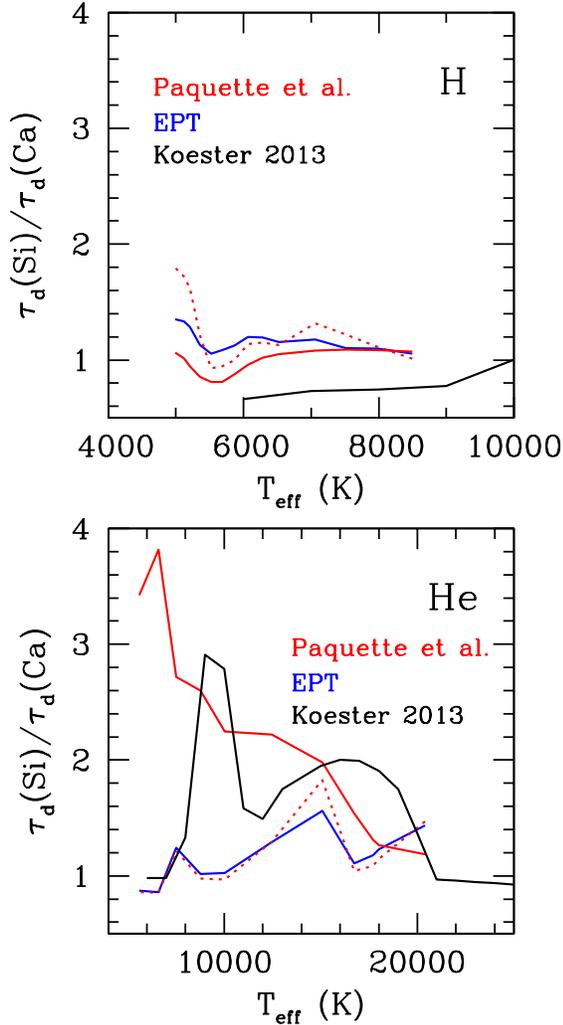}
   \caption{Ratio of the diffusion time scales of Si and Ca at the bottom of the convection zone for H (top panel) and
   He (bottom panel) envelopes.  The \cite{paquette86a} calculation is shown for ion charges from  the \cite{fbdt15} ionization model and the AA-TCP model
   with solid and dotted red lines, respectively (see Figures \ref{Zbar_in_H} and \ref{Zbar_in_He}). The ratios based on the AA-TCP model and 
   the EPT theory are shown in blue.  The calculation of \cite{koester13} is shown in black. Unlike in the preceding figures, the abscissa is 
    the effective temperature of a 0.6$\,$M$_\odot$ white dwarf that evolves to 
    the left as it cools (Tables \ref{tab:tab_cz_h} and \ref{tab:tab_cz_he}). The two panels are plotted on different $\teff$ scales.  See text for details.
   [{\it See the electronic edition of the Journal for a color version of this figure.}]}
    \label{tau_ratio}
\end{figure}

Figure \ref{tau_ratio} shows the ratio of diffusion time scales of Ca and Si in H and He white dwarf envelopes as a function of $\teff$ along the cooling sequences of Tables \ref{tab:tab_cz_h} 
and \ref{tab:tab_cz_he}.  For each case, we present the 1) Paquette et al. calculation,
2) the Paquette et al.  calculation but with the ion charges from the AA-TCP plasma model, 3) the EPT calculation and 4) the results of \citet{koester13}. 
For diffusion in H envelopes, we find that Ca and Si diffuse 
at approximately the same time scale for the range studied here ($\teff = 5000$ -- 8500$\,$K). Substituting the charges from the AA-TCP model in the Paquette et al. formalism increases the diffusion 
time scale of Si because of the larger ${\bar Z}$(Si) (Figure \ref{Zbar_in_H}). The effect becomes significant for the coolest H-rich stars, almost reaching a factor of two. Switching to the EPT calculation  
somewhat compensates for the higher charge of Si ions and the difference with Paquette et al. is $\lesssim 20$\%. The behavior in He envelopes is quite different. Here the ratio of diffusion time scales
from Paquette et al. ranges from $\sim 1.1$ at $\teff \sim 20000\,$K to greater than 3 for $\teff$ below 7500$\,$K. Applying the AA-TCP charges to the Paquette et al. calculation brings 
it in excellent agreement 
with the EPT calculation, which is generally well below the former result. The much larger charge of Si obtained with the AA-TCP model in the deep convection zones of low-$\teff$ envelopes accounts for the 
large drop in the time scale ratio.  Thus, in DZ (He-rich) stars with $\teff < 10000\,$K, the EPT theory (and the underlying AA-TCP plasma model) 
predicts a diffusion time scale ratio $\tau_{\rm d}$(Si)/$\tau_{\rm d}$(Ca) at the bottom of the convection zone that is about one third of that predicted with the Paquette et al. model.

We further compare the ratio of diffusion time scales with the results of \cite{koester09} as updated in \cite{koester13}. Both use the \cite{paquette86a} fits to the collision integrals.
Those are shown by the black lines in Figure \ref{tau_ratio}. A detailed 
discussion is not possible because the physical conditions at the bottom of the convection zone where the diffusion time scales are evaluated are not given in \cite{koester13}, but the pressures, 
temperatures and masses of the convection zone of
\cite{koester09} are close to those in Tables \ref{tab:tab_cz_h} and \ref{tab:tab_cz_he}. The ion charges are calculated with an independent ionization model and are not 
specified, however. For diffusion in hydrogen envelopes, $\tau_{\rm d}$(Si)/$\tau_{\rm d}$(Ca) is 
systematically lower than Paquette 
et al. and the EPT calculations by 40\% to 80\%. This probably arises from a combination of different stellar structures and ionization model.  More striking is the case of helium envelopes where 
large deviations between Paquette et al. and EPT are found, with considerable variation with $\teff$. The double hump structure seen in the \cite{koester13} curve is very likely due to the ionization model 
as the peaks occur approximately where changes in ionization due to electronic shells are expected.
This again highlights the importance of the ionization model in computing diffusion time scales.

In the cases of equilibrium and episodic accretion, a factor of $\sim 3$ change in the ratio of diffusion time scales leads to significant changes in the inferred composition 
of the accreted material.  The latter often departs from that measured in meteorites and solar system bodies and varies from star to star, leading to various
interpretations regarding other planetary systems and their evolution at very late stages of stellar evolution (see, for example, 
\cite{raddi15}, \cite{kawka16}, \cite{hollands18}, \cite{harrison18}, \cite{doyle19}). It would be prudent to consider that the inferred composition of the accreted material is affected by 
uncertainties in the diffusion coefficients that are not negligible, particularly in cool He-rich white dwarfs. 

\subsection{Other mixing processes}

The description of convection with the mixing length theory used here is rather simplistic. Two processes, thermohaline mixing and convective overshooting 
at the bottom of the convection zone, can increase the effective extent of mixing (i.e. $M_{\sss\rm CZ}$) considerably. The metal enrichment of the upper layers of a white dwarf that is accreting solid material 
causes an inverted molecular weight gradient that can trigger a thermohaline (double diffusive) instability \citep{deal13, wachlin17, bauer18, bauer19}. Because it is driven by the 
gradient of molecular weight,   the thermohaline instability is favored by higher accretion rates on white dwarfs with thin superficial convection zones where the accreted material is  
more concentrated.  \cite{bauer19} have shown that thermohaline mixing has a minimal effect in DA white dwarfs with $\teff \lesssim 9000\,$K and in DB white dwarfs below 18000$\,$K. Our 
study of diffusion coefficients focuses on white dwarfs near these limits and cooler (Tables \ref{tab:tab_cz_h} and \ref{tab:tab_cz_he}), thus
thermohaline mixing will not affect our results.  

Convective overshooting is caused by the momentum of downward moving fluid plumes that allows them to move past the convective/radiative boundary 
as defined by the Schwarzschild criterion in 1-dimensional stellar models. These fluid parcels are decelerated as they move in the convectively stable region but provide effective mixing well below the boundary  
calculated from the mixing length theory. Three-dimensional radiative hydrodynamics simulations of convection in white dwarfs indicate that the mixing due to overshooting
extends over a few pressure scale heights and increases the mass of the mixed region $M_{\sss\rm CZ}$ by up to 2.5 orders of magnitude \citep{tremblay15, kupka18, cunningham19}. 
Those studies have so far been limited to hydrogen envelopes and $\teff > 11400\,$K (for $\log g=8$), which is above the range of our study of 
diffusion coefficients.  Little is known about overshooting in the cooler hydrogen or helium envelopes relevant to this study. Nonetheless, we can make a general observation as to 
how overshooting bears on our results.

The net effect of convective overshooting is to increase the depth where metals diffuse out of the convection zone, increasing $M_{\sss\rm CZ}$. Everything else being equal, it 
implies a higher accretion rate for a given observed surface abundance, as suggested
by Equation (\ref{x2_equil}). \footnote{The actual dependence is not linear since $\tau_{\rm d}$ also increases with depth  \citep{kupka18, cunningham19}.} At deeper levels, the temperature,
density, as well as the plasma coupling and degeneracy will be higher than the values reported in Tables \ref{tab:tab_cz_h} and \ref{tab:tab_cz_he}. 
However, our white dwarf evolution sequences show that a two order of magnitude increase in $M_{\sss\rm CZ}$ results in an increase in temperature of $\Delta \log T \lesssim 0.5$ and of 
$\Delta \log \rho \lesssim 1.5$ at the bottom of the mixing region. 
The plasma coupling parameter of the background element (H or He) increases by a modest factor of 1 -- 1.4. At higher densities and 
temperatures the charge of the heavy ion  will be larger, increasing the ion-background coupling that primarily affects the diffusion coefficient. 
Taking overshooting into account, we expect the differences in the ratio $\tau_{\rm d}({\rm Si})/\tau_{\rm d}({\rm Ca})$
between models to be very similar if not larger than those shown in Figure \ref{tau_ratio}. 

\section{Conclusion}
\label{sec:conclusion}

We have revisited the calculation of diffusion coefficients using an advanced model for the partially ionized plasma found in the envelopes of cool white dwarfs.
This model combines an average atom model with the integral equations of fluid theory for a two-component plasma of classical ions and quantum electrons (the ``AA-TCP'' model). 
This plasma model describes
self-consistently the bound and free electronic states and the interactions between ions and electrons for any degree of plasma coupling and electron degeneracy. It accounts for temperature and 
pressure ionization equally well without introducing somewhat heuristic concepts such as continuum lowering or occupation probabilities.  The model solves for
the average charge of the ions, the interaction potentials and the correlation functions. The ionic pair distribution function can then be used in an extension of the Boltzmann equation called
Effective Potential Theory (``EPT'') to compute ionic transport coefficients, such as the inter-diffusion coefficient and the thermal diffusion factor. We have looked at 
the diffusion coefficients of Ca and Si ions at the bottom of the convection zone of cool white dwarfs with hydrogen and helium envelopes, and compared with the widely used coefficients of 
\cite{paquette86a}.

For the same set of conditions, we have calculated the coefficient of diffusion $D_{12}$ with the \cite{paquette86a} formalism as modified by \cite{fbdt15}, with the Effective Potential Theory 
and with classical molecular dynamics. The latter two methods are based on the same ion-ion pair potential obtained from the AA-TCP model and we find excellent 
agreement within the uncertainty and scatter of the molecular dynamics simulations, which further validates the EPT in this application.

We have shown that in weakly to moderately coupled plasmas, the \cite{paquette86a} calculation of $D_{12}$ is very good but becomes increasingly inaccurate as the plasma becomes more strongly 
coupled. In the context of stellar astrophysics, the \cite{paquette86a} approach is perfectly adequate for all normal stars where the plasma is weakly coupled, as well as for all DA white dwarfs and
DB white dwarfs with $\teff \gtrsim 15000\,$K. In cooler white dwarfs with helium envelopes, some of the assumptions underlying the \cite{paquette86a} coefficients become inadequate, 
resulting in diffusion coefficients that are underestimated by over a factor of two in the coolest models. The ionic thermal diffusion factor $\alpha_{12}$ is generally in good agreement  
between the two calculations
with the EPT value being $\sim 20$\% lower across all coupling regimes.

The diffusion coefficients of \cite{paquette86a}, \cite{fbdt15}, \cite{koester09} and \cite{stanton16} all based on the evaluation of the collision integrals that appear 
in the Chapman-Enskog solution or the resistance coefficients of Burgers,
with a static screened Coulomb (i.e. Yukawa) potential for the ion-ion interaction. However, they differ in how the ion charge is calculated and thus in their 
ion-ion  potentials, even though they share the same Yukawa functional form. The calculation of the ionization of a metal in white dwarf envelopes involves pressure ionization and is challenging, 
and the resulting uncertainty in the diffusion coefficients was acknowledged early on \citep{dupuis92}. 
We investigated this effect by computing $D_{12}$ with the \cite{paquette86a} model with the \cite{fbdt15} model of ionization and with the charges obtained with the AA-TCP 
model that has the most realistic microscopic plasma physics of the models considered here. We found that the average ionic charge of the heavy ion can vary substantially between models and that 
$D_{12}$ is affected at the $\sim 30$\% level. 
A further consideration, which we did not address, is that the heavy element will have a distribution of charge states,\footnote{The background species (H or He) is fully ionized at the bottom of
the convection zone.} each with a different $D_{12}$. 
In this case, a proper description of diffusion would be to treat each ionization stage as a separate species rather than the diffusion of an ion with an average charge 
\citep{dupuis92, koester09, bauer19}.

For simplicity and for illustration purposes, we considered diffusion time scales at the base of the convection zone as defined by the mixing length theory 
and the Schwarzschild stability criterion in 1-dimensional models. The thermohaline instability and convective overshooting are two processes that can extend vertical mixing to much greater depths. 
The thermohaline instability occurs only in white dwarfs that are hotter than those we considered here \citep{bauer18, bauer19}, i.e. in stars where the plasma is weakly coupled and 
the \cite{paquette86a} coefficients are reliable. Three-dimensional simulations of convective overshooting in white dwarfs show that the mixing zone can extend much deeper
than the nominal convection zone obtained with the mixing length theory. However, the corresponding increase in plasma coupling at the bottom of the mixing layer, and 
therefore the decrease in $D_{12}$, is modest, with little consequence on our results. 

Our results are of immediate relevance to the determination of the composition of accreted solid planetesimals from the observed 
abundance of metals in the atmospheres of white dwarfs. The elemental planetesimal composition inferred from the accretion/diffusion scenario provides important information as to the nature 
and origin of mature exoplanetary systems \citep{xu13, hollands18}.  The abundance of oxygen in particular 
has recently been shown to provide a measure of its fugacity in the solid accreted material which is an important clue to its geochemistry \citep{doyle19}.
This composition depends strongly on the relative diffusion time scales of the various elements. We found that the ratio
of diffusion time scales between Si and Ca changes by over a factor of three in cool DZ stars when applying a more sophisticated theory for the dense plasma and the calculation of diffusion 
coefficients than the widely used \cite{paquette86a} diffusion coefficients. 
In this study, we focused on two elements that are common in polluted white dwarfs, Si and Ca, which are somewhat similar with atomic numbers of 14 and 20, respectively. 
DZ stars show a much broader range of elements in their spectra, from C to Sr \citep{xu13}, and we expect that updated diffusion coefficients will deviate 
from Paquette et al. accordingly.  It would be of interest to extend this work to other heavy elements commonly observed in cool DZ stars, as well as to the diffusion of non-trace mixtures, such as
the inter-diffusion of H and He and of C and He.

To summarize, the model applied here combines a sophisticated plasma model and a modification of the Chapman-Enskog theory of transport in plasmas that provides a valuable compromise between 
physical realism and computational cost.  The \cite{paquette86a} formalism is perfectly suitable in weakly coupled plasmas and has the considerable advantage that the collision integrals 
can be calculated once and for all and scaled to any mixture of ions a posteriori.   Our approach requires a separate tabulation for each element pair or mixture. The accuracy gained
in the diffusion coefficients in cool white dwarfs and the astrophysical implications of the new  coefficients should motivate such an effort.
This initial application of the AA-TCP plasma model with the EPT theory to diffusion in white dwarfs demonstrates that advances in the modeling  of dense plasmas
can have important astrophysical consequences.

\acknowledgments
We thank the anonymous referee for carefully reading our manuscript and for a thoughtful report. This work was performed under the auspices of the U.S. Department of Energy under Contract No. 89233218CNA000001 and was supported in part by the
U.S. Department of Energy LDRD program at Los Alamos National Laboratory.

\newpage

\bibliographystyle{apj}
\bibliography{references}

\begin{thebibliography}{}

\bibitem[\protect\astroncite{Aller \& Chapman}{1960}]{ac60} Aller, L. H. \& Chapman, S. 1960, \apj, 132, 461
\bibitem[\protect\astroncite{Althaus \& Benvenuto}{2000}]{althaus00} Althaus, L. G. \& Benvenuto, O.J. 2000, \mnras, 317, 952
\bibitem[\protect\astroncite{Bastea}{2005}]{bastea05} Bastea, S. 2005, \pre, 71, 056405
\bibitem[\protect\astroncite{Baalrud \& Daligault}{2013}]{baalrud13} Baalrud, S.D. \& Daligault, J.  2013, \prl, 110, 235001  
\bibitem[\protect\astroncite{Baalrud \& Daligault}{2015}]{baalrud15} Baalrud, S.D. \& Daligault, J.  2015, \pre, 91, 063107   
\bibitem[\protect\astroncite{Baalrud \& Daligault}{2019}]{baalrud19} Baalrud, S.D. \& Daligault, J.  2019, Phys. Plasmas, 26, 082106
\bibitem[\protect\astroncite{Bauer \& Bildsten}{2018}]{bauer18} Bauer, E.B. \& Bildsten, L.  2018, \apjl, 859, 19
\bibitem[\protect\astroncite{Bauer \& Bildsten}{2019}]{bauer19} Bauer, E.B. \& Bildsten, L.  2019, \apj, 872, 96
\bibitem[\protect\astroncite{Beznogov \& Yakovlev}{2014}]{by14} Beznogov, M.V. \& Yakovlev, D.G. 2014, \pre 90, 033102        
\bibitem[\protect\astroncite{Blouin et al.}{2019}]{blouin19a} Blouin, S., Dufour, P., Thibeault, C., \& Allard, N. 2019, \apj, 878, 63  
\bibitem[\protect\astroncite{Brassard \& Fontaine}{2014}]{brassard14} Brassard, P. \& Fontaine, G.  2014, ASP Conf. Ser., 481, 221     
\bibitem[\protect\astroncite{Burakovzky et al.}{2013}]{burakovsky13} Burakovsky, L., Ticknor, C., Kress, J.D., Collins, L.A. \& Lambert, F. 2013, \pre 87, 023104 
\bibitem[\protect\astroncite{Burgers}{1969}]{burgers69} Burgers, J.M. 1969, Flow Equations for Composite Gases, (New York: Academic)
\bibitem[\protect\astroncite{Chapman \& Cowling}{1970}]{cc70} Chapman, S. \& Cowling, T.G. 1970 The mathematical theory of 
		             uniform gases (Cambridge: Cambridge University Press)
\bibitem[\protect\astroncite{Chayer et al.}{1995}]{chayer95} Chayer, P., Fontaine, G. \& Wesemael, F. 1995, \apjs, 99, 189 
\bibitem[\protect\astroncite{Cunningham et al.}{2019}]{cunningham19} Cunningham, T., Tremblay, P.-E., Freytag, B., Ludwig, H.-G. \& Koester, D. 2019, \mnras , 488, 2503
\bibitem[\protect\astroncite{Daligault}{2012}]{daligault12} Daligault, J. 2012, \prl 108, 225004
\bibitem[\protect\astroncite{Daligault et al.}{2016}]{daligault16} Daligault, J., Baalrud, S.D., Starrett, C.E., Saumon, D. \& Sjostrom, T. 2016, \prl 116, 075002   
\bibitem[\protect\astroncite{Danel, Kazandjian \& Z\'erah}{2012}]{danel12} Danel, J.-F., Kazandjian, L. \& Z\'erah, G. 2012, \pre 85, 066701
\bibitem[\protect\astroncite{Deal et al.}{2013}]{deal13} Deal, M., Deheuvels, S., Vauclair, G., Vauclair, S. \& Wachlin, F.C. 2013, \aap, 557, L12 
\bibitem[\protect\astroncite{Doyle et al.}{2019}]{doyle19} Doyle, A.E., Young, E.D., Klein, B., Zuckerman, B. \& , Schlichting, H. 2019, , Science, 366, 356 
\bibitem[\protect\astroncite{Dufour et al.}{2007}]{dufour07} Dufour, P., Bergeron, P., Liebert, J., Harris, H.C., Knapp, G.R., Anderson, S.F., Hall, 
	                     P.B., Strauss, M.A., Collinge, M.J. \& Edwards, M.C.  2007, \apj, 663, 1291
\bibitem[\protect\astroncite{Dupuis et al.}{1992}]{dupuis92} Dupuis, J., Fontaine, G., Pelletier, C. \& Wesemael, F. 1992, \apjs, 82, 505
\bibitem[\protect\astroncite{Dupuis et al.}{1993}]{dupuis93} Dupuis, J., Fontaine, G., Pelletier, C. \& Wesemael, F. 1993, \apjs, 84, 73
\bibitem[\protect\astroncite{Eliezer et al.}{2002}]{eliezer02} Elizier, S., Ghatak, A. \& Hora, H.  2002 Fundamentals of equations of state (World Scientific, New Jersey), Chap. 8
\bibitem[\protect\astroncite{Farihi}{2016}]{farihi16} Farihi, J. 2016, New Astron. Rev., 71, 1
\bibitem[\protect\astroncite{Feynman et al.}{1949}]{fmt49} Feynman, R.P., Metropolis, N. \& Teller, E., 1949, PhRv, 75, 1561
\bibitem[\protect\astroncite{Fontaine \& Michaud}{1979}]{fm79} Fontaine, G. \& Michaud, G., 1979, \apj 231, 826 
\bibitem[\protect\astroncite{Fontaine, Brassard \& Bergeron}{2001}]{fbb01} Fontaine, G., Brassard, P. \& Bergeron, P., 2001, \pasp 113, 409 
\bibitem[\protect\astroncite{Fontaine et al.}{2015}]{fbdt15} Fontaine, G., Brassard, P., Dufour, P. \& Tremblay, P.-E., 2015, ASP Conf. Ser. 493, 113 
\bibitem[\protect\astroncite{French et al.}{2012}]{french12} French, M. Becker, A., Lorenzen, W., Nettelmann, N. Bethkenhagen, M., Wicht, J. \& Redmer, R. 2012, \apjs, 202, 5
\bibitem[\protect\astroncite{Garc\'ia-Berro et al.}{2010}]{garcia-berro10} Garc\'ia-Berro, E., Torres, S., Althaus, L.G., 
	                     Renedo, I., Lor\'en-Aguilar, P., C\'orsico, A.H., Rohrmann, R.D., Salaris, M. \& Isern, J. 2010, 
	                     \nat 465, 194
\bibitem[\protect\astroncite{Grabowski et al.}{2020}]{graboske20} Grabowski, P. E., Hansen, S.B., Murillo, M.S. and 37 authors 2020, submitted to High Ener. Dens. Phys.
\bibitem[\protect\astroncite{Green}{1954}]{green54} Green, M. S. 2020, J. Chem. Phys. 22, 398
\bibitem[\protect\astroncite{Hansen \& McDonald}{2013}]{hmcd} Hansen, J.-P. \& McDonald, I. R. 2013, Theory of simple liquids,
			     Chap. 7 (Oxford: Academic Press) 
\bibitem[\protect\astroncite{Hansen, McDonald \& Pollock}{1975}]{hansen75} Hansen, J.-P., McDonald, I. R. \& Pollock, E.L. 1975, \pra 11, 1025 
\bibitem[\protect\astroncite{Hansen, McDonald \& Vieillefosse}{1975}]{hansen79} Hansen, J.-P., McDonald, I. R. \& Vieillefosse, P. 1979, \pra 20, 2590 
\bibitem[\protect\astroncite{Harrison et al.}{2018}]{harrison18} Harrison, J.H.D., Bonsor, A. \& Madhusudhan, N. 2018, \mnras 479, 3814
\bibitem[\protect\astroncite{Haxhimali et al.}{2014}]{haxhimali14} Haxhimali, T., Rudd, R., Cabot, W.H. \& Graziani, F. 2014, \pre, 90, 023104
\bibitem[\protect\astroncite{Hollands et al.}{2018}]{hollands18} Hollands, M.A., G\"ansicke, B.T. \& Koester, D. 2018, \mnras 477, 93
\bibitem[\protect\astroncite{Hummer \& Mihalas}{1988}]{hm88} Hummer, D. G. \& Mihalas, D. 1988, \apj 331, 794
\bibitem[\protect\astroncite{Iben \& McDonald}{1985}]{im85} Iben, I. Jr. \&  MacDonald, J. 1985, \apj 296, 540 
\bibitem[\protect\astroncite{Jakse \& Pasturel}{2013}]{jakse13} Jakse, N. \& Pasturel, A. 2013, Scientific Rep. 3, 3135 
\bibitem[\protect\astroncite{Jura}{2003}]{jura03} Jura, M. 2003, \apjl 594, 91
\bibitem[\protect\astroncite{Jura \& Young}{2014}]{jura14} Jura, M. \& Young, E.D. 2014, Ann. Rev. Earth Plan. Sci. 42, 45   
\bibitem[\protect\astroncite{Kawka \& Vennes}{2016}]{kawka16} Kawka, A. \& Vennes, S., 2016, \mnras, 458, 325
\bibitem[\protect\astroncite{Koester}{2009}]{koester09} Koester, D. 2009, \aap, 498, 517
\bibitem[\protect\astroncite{Koester}{2013}]{koester13} Koester, D. 2013, {\tt http://www1.astrophysik.uni-kiel.de/~koester/ \\ astrophysics/astrophysics.html}
\bibitem[\protect\astroncite{Koester et al.}{2014}]{koester14} Koester, D., G\"ansicke, B. T. and Fahiri, J. 2014, \aap, 566, 34
\bibitem[\protect\astroncite{Kress et al.}{2011}]{kress11} Kress, J.D., Cohen, J.S., Kilcrease, D.P., Horner, D.A. \& Collins, L.A. 2011, \pre, 83, 026404
\bibitem[\protect\astroncite{Kupka et al.}{2018}]{kupka18} Kupka, F., Zaussinger, F. \& Montgomery, M.H. 2018, \mnras, 474, 4660
\bibitem[\protect\astroncite{Kubo}{1957}]{kubo57} Kubo, R. 1957, J. Phys. Soc. Japan, 12, 570
\bibitem[\protect\astroncite{Lambert, Cl\'erouin \& Mazevet}{2006}]{lambert06} Lambert, F., Cl\'erouin, J. \& Mazevet, S. 2006, Europhys. Lett. 75, 681
\bibitem[\protect\astroncite{Liboff}{1959}]{liboff59} Liboff, R.L. 1959, Phys. Fluids, 2, 40
\bibitem[\protect\astroncite{Macquarrie}{1976}]{mcquarrie} McQuarrie, D. A. 1976, Statistical Mechanics, (Harper \& Row: New York) 
\bibitem[\protect\astroncite{Mason et al.}{1967}]{mason67} Mason, E.A., Munn, R.J. \& Smith, F.J.  1967, Phys. Fluids, 10, 1827
\bibitem[\protect\astroncite{Meyer et al.}{2015}]{meyer14} Meyer, E.R.,  Kress, J.D., Collins, L.A.  \& Ticknor, C. 2014, \pre, 90, 043101
\bibitem[\protect\astroncite{Michaud}{1970}]{michaud70} Michaud, G. 1970, \apj, 160, 641
\bibitem[\protect\astroncite{Michaud et al.}{1976}]{michaud76} Michaud, G., Charland, Y., Vauclair, S. \& Vauclair, G. 1976, \apj, 210, 447
\bibitem[\protect\astroncite{Muchmore}{1984}]{muchmore84} Muchmore, D.O. 1984, \apj, 278, 769
\bibitem[\protect\astroncite{Paquette et al.}{1986a}]{paquette86a} Paquette, C., Pelletier, C., Fontaine, G. \& Michaud, G., 1986a, \apjs, 61, 177
\bibitem[\protect\astroncite{Paquette et al.}{1986b}]{paquette86b} Paquette, C., Pelletier, C., Fontaine, G. \& Michaud, G., 1986b, \apjs, 61, 197
\bibitem[\protect\astroncite{Paxton et al.}{2015}]{paxton15} Paxton, B., Marchant, P., Schwab, J., Bauer, E.B., Bildsten, L., Cantiello, M.,
	                     Dessart, L., Farmer, R., Hu, H., Langer, N., Townsend, R.H.D., Townsley, D.M. \& Timmes, F.X. 2015, \apjs, 220, 15
\bibitem[\protect\astroncite{Pelletier et al.}{1986}]{pelletier86} Pelletier, C., Fontaine, G., Wesemael, F., Michaud, G. \& Wegner, G. 1986, \apj, 307, 242
\bibitem[\protect\astroncite{Raddi et al.}{2015}]{raddi15} Raddi, R., G\"ansicke, B.T., Koester, D., Farihi, J., Hermes, J.J., Scaringi, S., Breedt, E. \& Girven, J. 2015, \mnras, 450, 2083
\bibitem[\protect\astroncite{Rudd et al.}{2012}]{rudd12} Rudd, R.E., Cabot, W.H., Caspersen, K.J., Greenough, J.A., Richards, D.F., Streitz, F.H.  
	                     \& Miller, P.L. 2012, \pre, 85, 031202
\bibitem[\protect\astroncite{Salin \& Gilles}{2006}]{salin06} Salin G., \& Gilles, D. 2006, J. Phys. A: Math. Gen., 39, 4517
\bibitem[\protect\astroncite{Saumon \& Starrett}{2020}]{saumon20} Saumon, D., \& Starrett, C.E. 2020, in Shock Compression of Condensed Matter -- 2019,
	                     AIP Conference proceedings, ed. T. Germann, M. Lane, M. Armstrong, in press.
\bibitem[\protect\astroncite{Saumon et al.}{2014}]{saumon14} Saumon, D., Starrett, C.E., Anta, J.A., Daughton, W. \& Chabrier, G. 2014, in Frontiers and 
			     Challenges in Warm Dense Matter, Lecture Notes in Computational Science and Engineering 96, ed. F. Graziani, M.P. Desjarlais, R. Redmer, S.B. Trickey, 
			     (Springer: Switzerland) p151
\bibitem[\protect\astroncite{Schatzman}{1958}]{schatzman58} Schatzman, E. 1958, White Dwarfs, (Amsterdam: North-Holland)
\bibitem[\protect\astroncite{Shaffer et al.}{2017}]{shaffer17} Shaffer, N.R., Baalrud, S.D. \& Daligault J. 2017, \pre, 95, 013206
\bibitem[\protect\astroncite{Sjostrom \& Daligault}{2015}]{sjostrom15} Sjostrom, T. \& Daligault, J. 2015, \pre, 92, 063304
\bibitem[\protect\astroncite{Stanton \& Murillo}{2016}]{stanton16} Stanton, L.G. \& Murillo, M.S. 2016, \pre, 93, 043203
\bibitem[\protect\astroncite{Starrett \& Saumon}{2013a}]{starrett13} Starrett, C.E. \& Saumon, D. 2013a, \pre, 87, 013104
\bibitem[\protect\astroncite{Starrett \& Saumon}{2013b}]{starrett13err} Starrett, C.E. \& Saumon, D. 2013b, \pre, 88, 059901(E)   
\bibitem[\protect\astroncite{Starrett \& Saumon}{2014a}]{starrett_is} Starrett, C.E. \& Saumon, D. 2014a, High. Ener. Dens. Phys., 10, 35          
\bibitem[\protect\astroncite{Starrett \& Saumon}{2014b}]{starrett_mixtures} Starrett, C.E., Saumon, D., Daligault, J. \& Hamel, S.  2014b, \pre, 90, 033110  
\bibitem[\protect\astroncite{Starrett et al.}{2015a}]{starrett_pamd} Starrett, C.E., Daligault, J. \& Saumon , D. 2015a, \pre, 91, 033101                 
\bibitem[\protect\astroncite{Starrett \& Saumon}{2015b}]{starrett15_Al} Starrett, C.E. \& Saumon , D. 2015b, \pre, 92, 033101     
\bibitem[\protect\astroncite{Starrett \& Saumon}{2016}]{starrett_eos} Starrett, C.E. \& Saumon, D. 2016, \pre, 93, 063206       
\bibitem[\protect\astroncite{Ticknor et al.}{2015}]{ticknor15} Ticknor, C., Collins, L.A.  \& Kress, J.D. 2015, \pre, 92, 023101
\bibitem[\protect\astroncite{Ticknor et al.}{2016}]{ticknor16} Ticknor, C., Kress, J.D., Collins, L.A., Cl\'erouin, J., Arnault, P. \& Decoster, A. 2016, \pre, 93, 063208 
\bibitem[\protect\astroncite{Tremblay et al.}{2015}]{tremblay15} Tremblay, P.-E., Ludwig, H.-G., Freytag, B., Fontaine, G., Steffen, M. \& Brassrad, P. 2015, \apj, 799, 142
\bibitem[\protect\astroncite{Wachlin et al.}{2017}]{wachlin17} Wachlin, F.C., Vauclair, G., Vauclair, S., \& Althaus, L.G. 2017, \aap, 601, A13
\bibitem[\protect\astroncite{Xu et al.}{2013}]{xu13} Xu, S., Jura, M., Klein, B., Koester, D. \& Zuckerman, B. 2013, \apj, 766, 132
\bibitem[\protect\astroncite{Zeidler-K. T. et al.}{1986}]{zeidler86} Zeidler-K. T., E.M., Weidemann, V. \& Koester, D. 1986, \aap, 155, 356

\end{thebibliography}

\clearpage


\end{document}